\documentclass[acmsmall, nonacm]{acmart}

\usepackage{algorithm,algorithmic}
\usepackage{subfig}
\usepackage{graphicx}
\AtBeginDocument{%
  \providecommand\BibTeX{{%
    \normalfont B\kern-0.5em{\scshape i\kern-0.25em b}\kern-0.8em\TeX}}}

\setcopyright{acmcopyright}
\copyrightyear{2024}
\acmYear{2024}

\acmConference[ACM 2024]{ACM Journal on Emerging Technologies in Computing Systems}{}{}
\acmBooktitle{ACM Journal on Emerging Technologies in Computing Systems}
\acmPrice{15.00}
\acmISBN{978-1-4503-XXXX-X/18/06}

\begin{document}

\title{Leveraging Error Resilience of Iterative Algorithms for Energy Efficiency: from Concept to Implementation}
\titlenote {This paper is an extended version of \cite{g1_analysis} and \cite{g2_accelerator}.}

\author{G.A. Gillani}

\email{s.ghayoor.gillani@utwente.nl}
\affiliation{%
  \institution{CAES Group, University of Twente}
}

\author{A. Krapukhin}
\authornote{A. Krapukhin is currently with AIRI. However, his contributions to this paper are from the year when he was a graduate student at the University of Twente from (Jan, 2018) to (Jan, 2019). Therefore, this work has no collaboration with AIRI. }
\affiliation{%
  \institution{CAES Group, University of Twente}
  }
\email{larst@affiliation.org}


\author{A.B.J. Kokkeler}
\affiliation{\institution{Radio Systems Group, University of Twente}}

\renewcommand{\shortauthors}{G.A. Gillani, et al.}

\begin{abstract}
  \textbf{Iterative algorithms} are widely used in digital signal processing
applications. With the case study of radio astronomy calibration processing,
this work contributes towards revealing and exploiting the intrinsic error
resilience of iterative algorithms for energy efficiency benefits. We consider iterative
methods that use a convergence criterion as a quality metric to terminate the
iterative computations. We propose an adaptive statistical approximation
model for high-level resilience analysis that provides an opportunity to divide
an iterative algorithm into exact and approximate iterations. We realize
an energy-efficient accelerator based on a heterogeneous architecture, where
the heterogeneity is introduced using accurate and approximate processing
cores. Our proposed methodology exploits the error-resilience of the algorithm,
where initial iterations are processed on approximate modules while the later
ones on accurate modules. The proposed accelerator design does not increase
the number of iterations as compared to that of an accurate counterpart and
provides sufficient precision to converge to an acceptable solution. Our implementation using TSMC 40nm Low Power (TCBN40LP) technology shows 23\% savings in electrical energy consumption.
\end{abstract}

\ccsdesc[500]{Computing methodologies~Model development and analysis}
\ccsdesc[300]{Hardware~Power and energy}
\ccsdesc[100]{Computer systems organization~Heterogeneous (hybrid) systems}


\keywords{approximate computing, iterative algorithms, energy efficiency, accelerators, radio astronomy}

\begin{teaserfigure}
    \centering
\includegraphics[width=1\textwidth]{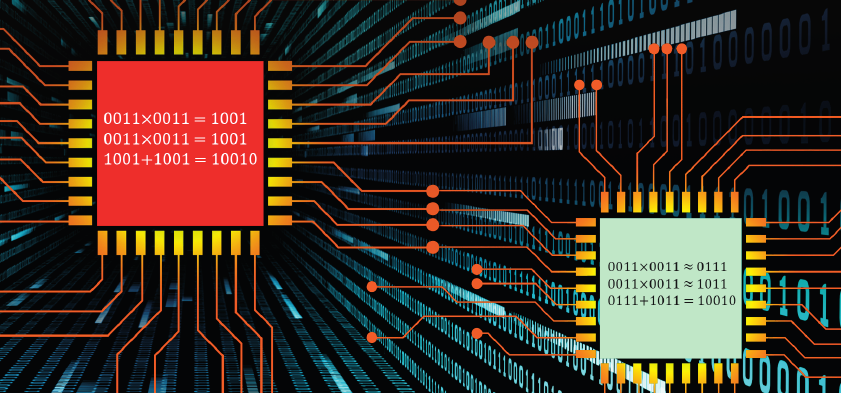}
    \captionsetup{labelformat=empty}
    \caption{Accurate chips are hot (energy-hungry), approximate chips are cool but imprecise; their combination is just right!}
    \Description{figure description}
  \end{teaserfigure}

\maketitle

\section{Introduction}
Increasing hardware efficiency is one of the major targets to innovate computing
devices. This includes the following, (1) reducing the size/chip-area of a transistor, i.e., increasing the number of transistors per unit area (transistor density),
(2) reducing the power consumption of a transistor to keep the power density
constant while the transistor density is increased, (3) increasing the speed, i.e.,
increasing the performance. The increase in hardware efficiency is generally
achieved by the advancements in Very Large Scale Integration (VLSI) technology. The improvements in transistor density are more or less following Moore’s
law \cite{84_moore}. The law states that the transistor density doubles every 1.5 years. For
that matter, we have been witnessing smaller sizes of devices that have gradually
brought gadgets in our hands. In the last century, the advancements in VLSI
technology were also following Dennard’s scaling of keeping the power density
constant \cite{28_dennard}.

However, there are physical limitations to the increase in efficiency of computing
devices \cite{10_fundamental,73_fundamental}. One of the biggest challenges faced by designers today is power-
/energy-consumption (Dennard’s scaling) \cite{34_PowerChallenges}. The power density is not scaling
as well as compared to the transistor density \cite{13_fundamental,73_fundamental}. The consequence is that a part of an integrated circuit (IC) has to be turned-off to control the power
budget, bringing us to the era of \textit{dark silicon} \cite{33_dark, 73_fundamental, 113_dark}. While architectural
power management techniques like Dynamic Voltage and Frequency Scaling
(DVFS) and clock-/power-gating are not enough to meet the power challenges
\cite{115_crosslayer}, new computing paradigms have to be explored. One of the paradigm shifts
is to move from conventional ’always correct’ processing to processing where
controlled errors are allowed. Computing techniques that are based
on the latter paradigm are called \textit{approximate computing techniques} or in short
\textit{approximate computing}. 

Approximate computing can be regarded as an aggressive optimization because
it allows controlled inexactness and provides results with the bare minimum
accuracy to increase computing efficiency. An increase in computing efficiency
or simply \textit{efficiency} means reduction in computing costs like run-time, chip-area,
and power/energy consumption. The introduction of inexactness brings errors
in the intermediate and/or the final outcomes of the processing compromising
output quality, or simply the \textit{quality} of processing. Approximate computing
has shown high-efficiency gains for error-resilient applications like multimedia
processing, machine learning and search engines \cite{132_survey, own_survey}. Such applications
tolerate a quantified error within the computation while producing an acceptable
output.

Assessing error resilience inherent to an algorithm provides application-specific
insights towards approximate computing strategies. The error resilience analysis
is performed by applying approximations while monitoring the quality function
to verify that the approximations produce acceptable results. In this way,
approximate computing techniques are substantiated for an algorithm that can
be employed in the implementation phase to achieve the desired benefits. With the advent of accuracy-configurable architectures like adaptive voltage overscaling \cite{61_config} and adaptive processors comprised of configurable adder/multiplier blocks \cite{115_crosslayer}, where the quality-cost trade-off can be controlled at run time, it is
of remarkable importance to analyse an algorithm for Adaptive Error Resilience
(AER) as well. AER has the potential to reveal approximation opportunities even
in strict quality function (relatively less error-resilient) algorithms, where they
can be processed adaptively for varying approximation levels to gain hardware
efficiency benefits.

Moreover, heterogeneous architectures have the ability to handle various workloads/
algorithms efficiently while using different power and performance tradeoff
computing nodes, e.g., ARM’s big.LITTLE architecture \cite{54_biglittle}. In the context
of approximate computing, the definition of a heterogeneous architecture
extends further to include exact and inexact computing units, where control
instructions and sensitive computational parts run at precise cores while the
error-resilient parts run at the error-prone cores to achieve the overall efficiency
in speed and energy \cite{56_accordion, 108_thesis}.

Iterative methods\footnote{In this paper, the following terms mean the same: \textit{iterative methods}, \textit{iterative workloads}, and \textit{iterative algorithms}.} are common candidates of approximate computing like K-means \cite{76_best}, GLVQ training \cite{26_chippa}, and model predictive control \cite{104_flops}. Therefore,
we propose an Adaptive Statistical Approximation Model (Adaptive-SAM) to
perform the high-level error resilience analysis of iterative algorithms. The high-level
error resilience analysis refers to applying the statistical error based on a
Gaussian distribution and assessing the output-quality for a relaxed quality function \cite{26_chippa}. Adaptive-SAM performs better than the Statistical
Approximation Model (SAM) by quantifying the number of approximate
iterations (Nax) in addition to the statistical parameters: error mean (EM), error predictability (EP), and error rate (ER). Therefore, it can better exploit the heterogeneous/accuracy configurable
architectures by assigning resilient iterations to the approximate
computing cores/modes and the sensitive iterations to the exact counterparts.

Being pivotal building blocks of DSP architectures, approximate multipliers and
adders have been extensively researched for increased hardware efficiency \cite{50_review,kokkeler2024modeling,ullah2021high, ullah2022appaxo, farahmand2023design}. However, approximate accelerator designs for relatively bigger
algorithms have been of less attention yet. The Least Squares (LS) algorithm is
widely utilized in digital signal processing applications like image reconstruction
in radio astronomy \cite{86_naghibzadeh,107_stefcal}, medical \cite{89_fmri}, and synthetic aperture radar \cite{20_sar}
domains. Despite its importance, no approximate least squares accelerator design
has been investigated to the best of our knowledge.

Modern radio telescopes like Square Kilometer Array
(SKA) require highly energy-efficient processing architectures to process terabytes
of raw data per second. For instance, double-precision fused multiply-add
operations will require 7.2MW of power consumption in the medium-frequency
array of SKA if contemporary technology would be used \cite{51_ska}. In this regard, we
investigate an energy-efficient LS accelerator architecture based on a case study
of radio astronomy calibration processing. The aforesaid processing employs an iterative LS algorithm to compute sensors’ gains for a certain configuration of a
radio telescope. Although a single case study is utilized to illustrate the concept and to quantify the benefits, the given methodology provides an energy-efficient way to process iterative algorithms that utilize a convergence criterion to conclude their processing, see Section \ref{Adaptive_SAM} (Algorithm \ref{algo:eranalysis}).

The primary contribution of this work is an effective way of exploiting the error
resilience of iterative algorithms for achieving energy efficiency benefits. In this
regard, we present an energy-efficient heterogeneous architecture from its concept to implementation. Specifically, the following is presented in this paper,

» An adaptive statistical approximation model (Adaptive-SAM) for error
resilience analysis of iterative algorithms (Section \ref{Adaptive_SAM}).

» Error resilience analysis of a radio astronomy calibration algorithm (case
study) by performing the state-of-the-art statistical approximation model
(SAM) analysis and our proposed Adaptive-SAM analysis (Section \ref{s:eranalysis:Results}).

» An assessment of utilizing convergence criterion as a quality metric for
the error resilience analysis of iterative algorithms (Section \ref{s:eranalysis:Quality_Function} ). We
demonstrate the importance of quality function reconsideration for convergence
based iterative processes as the original quality function (the
convergence criterion) is not necessarily sufficient in the error resilience
analysis phase.

» An energy-efficient heterogeneous architecture for iterative algorithms
with a case study of an approximate least squares (LS) accelerator design
for radio astronomy calibration processing (Section \ref{eeAcceleratorDesign}).

\section{Background}
\label{s:eranalysis:relatedwork}
This section provides terminology and reviews some state-of-the-art approximate computing architectures that motivate statistical and adaptive-statistical error resilience analysis. Moreover, we discuss contemporary analysis methodologies/tools and the need of Adaptive-SAM analysis.
\subsection{Terminology}
\subsubsection{Efficiency}
In computing systems, the term \textit{computing efficiency} (or simply \textit{efficiency}) is used
in contradiction of computing resource usage or computing costs for executing a
specific task. It is defined as the output of a computing system per unit resource
input, e.g., energy efficiency of a floating-point processor is defined as floating-point operations per Joule or floating-point operations per second per Watt.
An increase in efficiency is referred to as a reduction in computing costs like
chip-area, runtime, and/or power/energy consumption. In this paper, an increase in computing efficiency or a decrease in computing
costs means the same; and a decrease in computing efficiency or an increase in
computing costs means the same.

\subsubsection{Quality}
The term \textit{output quality} (or simply \textit{quality}) is defined in contradiction to deviation from exact behavior or error. In this paper, unless explicitly mentioned, the
terms exact and accurate refer to a specified precision where there is no approximation involved with reference to the specified precision. For instance, if the
specified precision is 8-bit, the 8-bit design (e.g., an 8-bit multiplier) is considered as accurate or exact design. Any approximations, e.g., in terms of data (e.g.reducing the precision of inputs to 7-bit) or circuit (e.g., removing parts of 8-bit
multiplier circuit), brings an inexact or in-accurate or approximate entity, where
the entity is referred to as circuit and/or data. Moreover, an increase in quality is
referred to as a reduction in error. In literature, both terms (quality and error)
have been used to indicate the output quality. Also in this paper, an increase in
output quality or a decrease in output error means the same; and a decrease in
output quality or an increase in output error means the same.

\subsubsection{Accuracy and Precision}
Accuracy defines how close the output of a system is to that of the exact behavior.
We define exact behavior as the theoretical behavior of a computing
system for a specified precision. For instance, an 8-bit multiplier has an exact
behavior when 8-bit inputs are multiplied without any approximation. On the
other hand, precision is referred to as the amount of detail utilized in representing
the output \cite{2_ieeeGlossary}. Therefore, it provides a measure of how close the outputs of a
specific system are to each other. In the context of iterative algorithms, where
the iteration process is terminated based on convergence, the precision of an
approximate computing system is also important. In \textbf{Section 3}, we elaborate on
this difference based on our analysis of an iterative algorithm.

\subsection{Adaptive Accuracy Techniques}

Approximate computing utilizing adaptive accuracy techniques have been discussed in \cite{115_crosslayer} and \cite{61_config}. The adaptive voltage over-scaling (AVOS) \cite{61_config} has shown improvements in power efficiency ($25$\% to $30$\%) at a negligible quality loss for texture decompression application. The aforesaid scheme reduces the supply voltage until a specific number of errors are introduced and can increase the voltage again to attain error-free operation. Therefore, AVOS can adjust the quality-cost trade-off during run time. 

The idea of accuracy-configurable architectures composed of approximate accelerators has been proposed in \cite{115_crosslayer}. These architectures contain accuracy-configurable operators, such as multipliers and adders, which can change the computation mode from accurate to approximate and vice-versa during the run-time. This helps in run-time adjustment of the quality-cost trade-off based on the algorithm's error resilience. In this context, we argue that the aforementioned adaptive accuracy architectures can be better exploited provided that the error resilience profile of an iterative algorithm quantifies the number of approximate iterations in addition to the insights of promising approximations. This allows an adaptive employment of approximations during the run-time to gain target benefits.

\subsection{Error Resilience Analysis Techniques}
 
Based on literature study, here we summarize error resilience analysis techniques in the context of our proposed technique. The Quality of service (QoS) profiling utilizes loop perforation as a compiler pass in order to identify sub-computations that can be replaced with less accurate counterparts \cite{loopQos}. Intel's open-source approximate computing toolkit (iACT) assesses the scope of approximations within the applications using programmer annotated pragmas to analyse the programmer guided parts of the code \cite{iACT}. The iACT has the ability to apply static approximate transformations such as precision scaling and run-time approximations like memoization during the error resilience analysis. 

Automatic sensitivity analysis for approximate computing (ASAC) applies perturbation of program data to study the overall effects on the output quality \cite{asac}. Program analysis for approximation-aware compilation (PAC) provides a relatively faster way to study the accuracy requirement of each component in an algorithm \cite{pac}. Approximate C compiler for energy and performance trade-offs (ACCEPT) is another open-source tool that applies a conservative approach to perform safe approximate relaxation analysis within an algorithm \cite{108_thesis}. The aforesaid tools are limited in assessing the error-resilience of an algorithm as they do not cover all the approximation strategies and they do not provide a statistical error resilience profile to reduce the available design space (alternatives).

On the other hand, the application resilience characterization (ARC) framework \cite{26_chippa} includes a statistically distributed error injection model to generate the statistical profile of an algorithm. The aforesaid model is known as Statistical Approximation Model (SAM) and is utilized to perform the so-called high-level error resilience analysis. However, in order to better utilize the adaptive accuracy architectures, we need Adaptive-SAM analysis of iterative algorithms that can quantify the adaptive resilience by identifying the number of approximate iterations in addition to the statistical parameters.

\section{Error Resilience Analysis of Iterative Algorithms}
In this section, we elaborate on the Adaptive-SAM analysis methodology and present its significance with a case study of the radio astronomy calibration application. In practice, the error resilience of an application is quantified by injecting the errors defined by the approximation models (statistical or technique-specific) and monitoring the overall output of the application in compliance with the quality function. The range of error injection within an approximation model for which the quality function is satisfied can be regarded as the \textit{approximation space} of an application. Therefore, defining a quality function is a very deliberate task in the approximate computing domain. In Section \ref{s:eranalysis:Quality_Function}, we demonstrate that the original (precision-based) quality function of an iterative process is not necessarily sufficient in the error resilience analysis procedure, which requires defining an additional (accuracy-based) quality function to serve the purpose.

\subsection{Adaptive Statistical Approximation Model (Adaptive-SAM)} 
\label{Adaptive_SAM}
As discussed earlier, our aim is to improve the high-level error resilience analysis of iterative algorithms to better exploit accuracy configurable and heterogeneous architectures. In this regard, we present Adaptive-SAM that can replace SAM in the error resilience analysis methodology to provide the number of approximate iterations ($N_{ax}$) in addition to statistical parameters of the approximation space (EM, EP and ER). 

To elaborate on the Adaptive-SAM methodology, consider a template of an iterative algorithm shown in Algorithm \ref{algo:eranalysis}. The algorithm has inputs $(x_1, x_2, x_3, ...)$ and an output $(\text{K\_op})$. It iterates to improve the output by utilizing the inputs and previously computed result (last iteration), where $i$ is the current iteration and $N$ is the maximum number of iterations. The algorithm also uses an intermediate variable $(\text{im\_var})$ that is computed by calling a kernel: function\_im, having input arguments as $x_1, x_2, x_3, ...$ and the computed output of the previous iteration $\text{K\_op}(i-1)$. Then the output is computed by calling another kernel: function\_Kop that has input arguments $x_1, x_2, x_3, ...$ and $\text{im\_var}$. 

Subsequently, the convergence metric $(\text{convergence\_met})$ is computed by calling function\_conv kernel that considers the current output $\text{K\_op}(i)$ and the previously computed output $\text{K\_op}(i-1)$. Iterative algorithms may use different arithmetic operations within function\_conv, but the aim is generally to compute the improvement in result within two consecutive iterations. Finally, Algorithm \ref{algo:eranalysis} checks the convergence metric for the allowed tolerance limit $(\text{tol})$ based on the quality function of the iterative algorithm. If the convergence is reached, the iterative process is terminated to provide the final outcome.

\begin{algorithm}[t!]
 \caption{An Iterative Algorithm Template.}
\label{algo:eranalysis}
 \begin{algorithmic}[1]
 \renewcommand{\algorithmicrequire}{\textbf{Input:}}
 \renewcommand{\algorithmicensure}{\textbf{Output:}}
 \REQUIRE $x_{1},x_{2},x_{3},...$ 
 \ENSURE  $\text{K\_op}$
  \STATE Initialize $\text{K\_op}(0)$
  \FOR {$i = 1,2,...,N$}
  \STATE $\text{im\_var}= \text{function\_im}(x_1,x_2,x_3,...,\text{K\_op}(i-1));$
  \STATE $\text{K\_op}(i)= \text{function\_Kop}(\text{im\_var},x_1,x_2,x_3,...);$
  \STATE $\text{convergence\_met}= \text{function\_conv}(\text{K\_op}(i),\text{K\_op}(i-1));$
  
  \IF {($\text{convergence\_met} \le \text{tol}$)}
  \STATE break; \qquad // convergence reached
  \ENDIF
  \ENDFOR
 \end{algorithmic} 
 \end{algorithm}

As discussed in \cite{26_chippa}, the error resilience methodology identifies the dominant kernels in the first place. These kernels are selected based on the percentage of their run-time or floating point operations (FLOPs) relative to the overall algorithm. The kernels that have the higher share are regarded as dominant kernels as it is likely to attain desired benefits (area, power or latency) while approximating them. In Algorithm \ref{algo:eranalysis}, we assume that function\_Kop is a dominant kernel to explain Adaptive-SAM analysis. 

\begin{figure}[t!]
\centering
\includegraphics[width=4.4in]{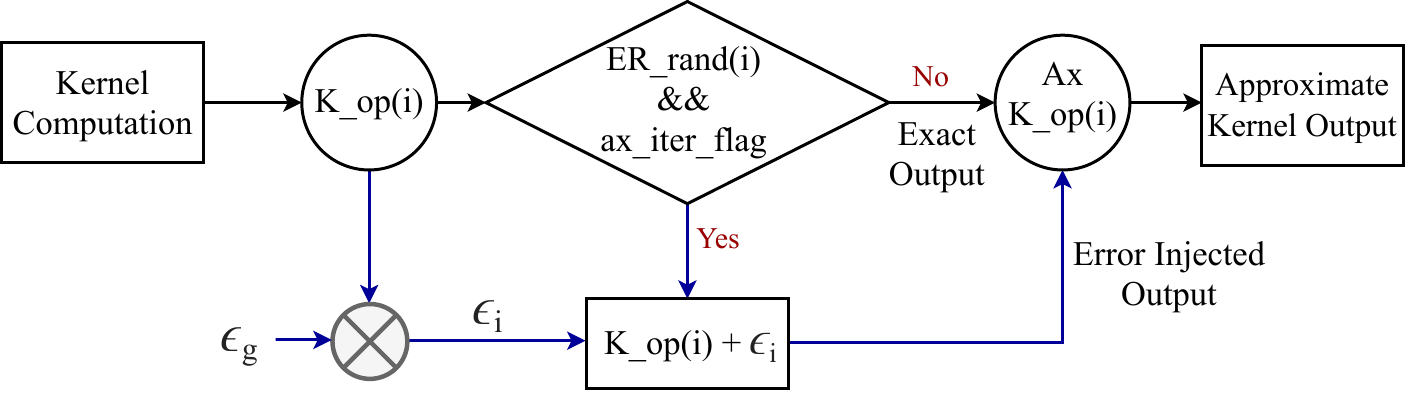}
\caption{Proposed high-level error resilience analysis model -- Adaptive-SAM. Our proposed model injects an error $\epsilon_\text{i}$ to an iteration of the dominant kernel $\text{K\_op}(i)$ if the randomized error and the approximate iterations flag ($\text{ER\_rand}(i)$ \&\& $\text{ax\_iter\_flag}$) allow error injection to the iteration.}
\label{fig:era2}
\end{figure}

Fig. \ref{fig:era2} shows the signal flow of Adaptive-SAM. We assume $N$ number of iterations where $i$ is the current iteration ($1 \leq i \leq N$). The output of iteration $i$  of the dominant kernel ($\text{K\_op}(i)$) is added to an error ($\epsilon_\text{i}$) if the randomized error ($\text{ER\_rand}(i)$) and approximate iterations flag ($\text{ax\_iter\_flag}$) allow error injection to this stage. This flag is controlled via a parameter: $N_{ax}$, which is the number of initial iterations to be instrumented with errors. It is important to note that $\epsilon_\text{i}$ is a Gaussian randomness that is generated in a relative manner, i.e., relative to the kernel output ($\text{K\_op}(i)$). This ensures that unsuitably big/small magnitudes of errors are not inserted into the kernel output. Also note that $\epsilon_\text{g}$ is a function of the EP and EM as given in Eq. \eqref{eq:epsilonG},
\begin{equation}
  \epsilon_\text{g} = \frac{(\text{EP} \times \text{randn}(i) + \text{EM})} {100}
\label{eq:epsilonG}
\end{equation}
where $\text{randn}(i)$ generates a random number with the standard normal distribution ($\mu= 0$ and $\sigma=1$). As an example, let $\text{EM}=10$ and $\text{EP}=0.2$, the probability is approximately >$99.7\%$ that $\epsilon_\text{i}$ lies between $9\%$ to $11\%$ of the $\text{K\_op}(i)$.
The approximate kernel output is assessed for quality function compliance to test the validity of the approximation space. Therefore, Adaptive-SAM quantifies the high-level error resilience of an application based on the acceptable range of a Gaussian randomness for the quantified approximate iterations. This helps to assign the number of approximate iterations to the inexact cores (fast and/or energy efficient) with the specified approximation space while running the exact iterations on the exact cores to better exploit the heterogeneous architectures. Similarly, in case of accuracy configurable architectures, the approximate iterations can run in approximate (error prone) hardware modes at higher energy efficiency and/or speed. 

\subsection{High-level Error Resilience Analysis} 
\label{s:eranalysis:Results}

We have applied SAM and Adaptive-SAM on an iterative algorithm, which is the radio astronomy calibration algorithm (StEFCal) \cite{107_stefcal}. The said algorithm has been instrumented with the run-time statistical error injection in Matlab. The quality function compliance has also been checked at run-time to determine the statistical error tolerance limits. In this way, we have quantified the high-level error resilience for StEFCal by using SAM and Adaptive-SAM models.   

\subsubsection{Radio Astronomy Calibration Processing}
\label{ss:era:raprocessing}
In radio telescopes, gain calibration improves the quality of sky images and reinforces the signal processing techniques against interference and spatial effects \cite{boonstra2003gain}. Calibration processing computes complex antenna gains in a radio telescope. The gains are estimated by minimizing the following \cite{107_stefcal},
\begin{equation}
\label{eq:difference}
  \|\textbf{V}-\textbf{GMG}^H\|^2_F
\end{equation}
where $\textbf{M}$ and $\textbf{V}$ represent the model and the measured visibilities, respectively ($\textbf{M}$, $\textbf{V}$ $\in\mathbb{C}$). $\textbf{G} = \text{diag}(\textbf{g})$ represents complex antenna gains ($\textbf{G}$ $\in\mathbb{C}$). In this work, we assume $124$ antennas.
 Therefore, the gains are $124$ complex numbers, so the matrix $\textbf{G}$ has a size of $124\times 124$ with gains on the diagonal and zeros at other entries. 

The calibration algorithm, also known as StEFCal (statistically efficient and fast calibration) \cite{107_stefcal}, is a strict quality-of-service method that estimates complex antenna gains $(g_p)$ for the $P$ sensors in a radio telescope. The algorithm computes a $\textbf{g}$ vector (representing $P$ gains) based on a measured signal/array covariance matrix ($\textbf{V}$) and the model covariance matrix ($\textbf{M}$), where each iteration $(i)$ computes $P$ independent linear least squares problems,
\begin{equation}
\label{eq:eranalysis:gpcomputation}
  g_p^{[i]}=\frac{\textbf{V}_{:,p}^H.\textbf{Z}_{:,p}^{[i-1]}} {(\textbf{Z}_{:,p}^{[i-1]})^H.\textbf{Z}_{:,p}^{[i-1]}}
\end{equation}
where $\textbf{V}_{:,p}^H$ is the hermitian transpose of array covariance matrix's $p\textit{th}$ column. $\textbf{Z}_{:,p}^{[i-1]}$ is the element-wise product of $\textbf{g}^{[i-1]}$ and the model covariance matrix's $p\textit{th}$ column ($\textbf{M}_{:,p}$),
\begin{equation}
\label{eq:eranalysis:zcomputation}
  \textbf{Z}_{:,p}^{[i-1]} =  (\textbf{M}_{:,p} \odot \textbf{g}^{[i-1]})
\end{equation}

It should be noted that $\textbf{g}^{[i-1]}$ is the antenna gains vector computed in the previous iteration. In our experiments, representative input data ($\textbf{V}$ and $\textbf{M}$ matrices) of the LOFAR facility \cite{LOFAR} has been utilized (for $P=124$). 

The convergence criterion of StEFCal is based on the relative length of the difference of consecutive iterations' solution vectors in the Euclidean space \cite{107_stefcal},
\begin{equation}
  \text{Convergence}= \frac{\|\textbf{g}^{[i]} - \textbf{g}^{[i-1]}\|_F} {\|\textbf{g}^{[i]}\|_F} \leq 1.10^{-6}
\label{eq:converg}
\end{equation}
In our initial experiments, we defined our quality function solely based on the convergence criterion. However, this proved to be insufficient in the error resilience analysis process as will be explained in Section \ref{s:eranalysis:Quality_Function}. Therefore, we have defined an additional quality metric: Diff\_rel, which is the relative difference in length between the exact (ex) and approximate (ax) solution vectors, 

\begin{equation}
  \text{Diff\_rel}=\frac{\|\textbf{g}_{ex}^{[i]}-\textbf{g}_{ax}^{[i]}\|_F} {\|\textbf{g}_{ex}^{[i]}\|_F} \leq 1.10^{-5}
\label{eq:diff_rel}
\end{equation}

We assume that the \textit{quality acceptance} is achieved, if and only if both the convergence (Eq. \ref{eq:converg}) and Diff\_rel (Eq. \ref{eq:diff_rel}) criteria are satisfied.

\subsubsection{Simulation Results of SAM and Adaptive-SAM}
To apply SAM and Adaptive-SAM, the initial steps are the following, distinguishing dominant kernels that run at least $1$\% of the total execution time and identifying error resilience by injecting the random errors in the outputs of the dominant kernels. 
Three dominant kernels have been identified in StEFCal by using profiling \cite{FLOPS}. One of them is $\textbf{Z}$ computation (Eq. \ref{eq:eranalysis:zcomputation}), which brings $27$\% of the computational load. The other two are the dot products (Eq. \ref{eq:eranalysis:gpcomputation}), which bring $72$\% of the computational load. We have analysed the aforesaid three kernels for high-level error resilience. As the response of the dot products is almost similar to $\textbf{Z}$ computation, we only present the simulation results for $\textbf{Z}$ computation here. They are sufficient to demonstrate the comparison between SAM and Adaptive-SAM outcomes. 

\begin{figure}[t!]
\centering
\includegraphics[width=3in]{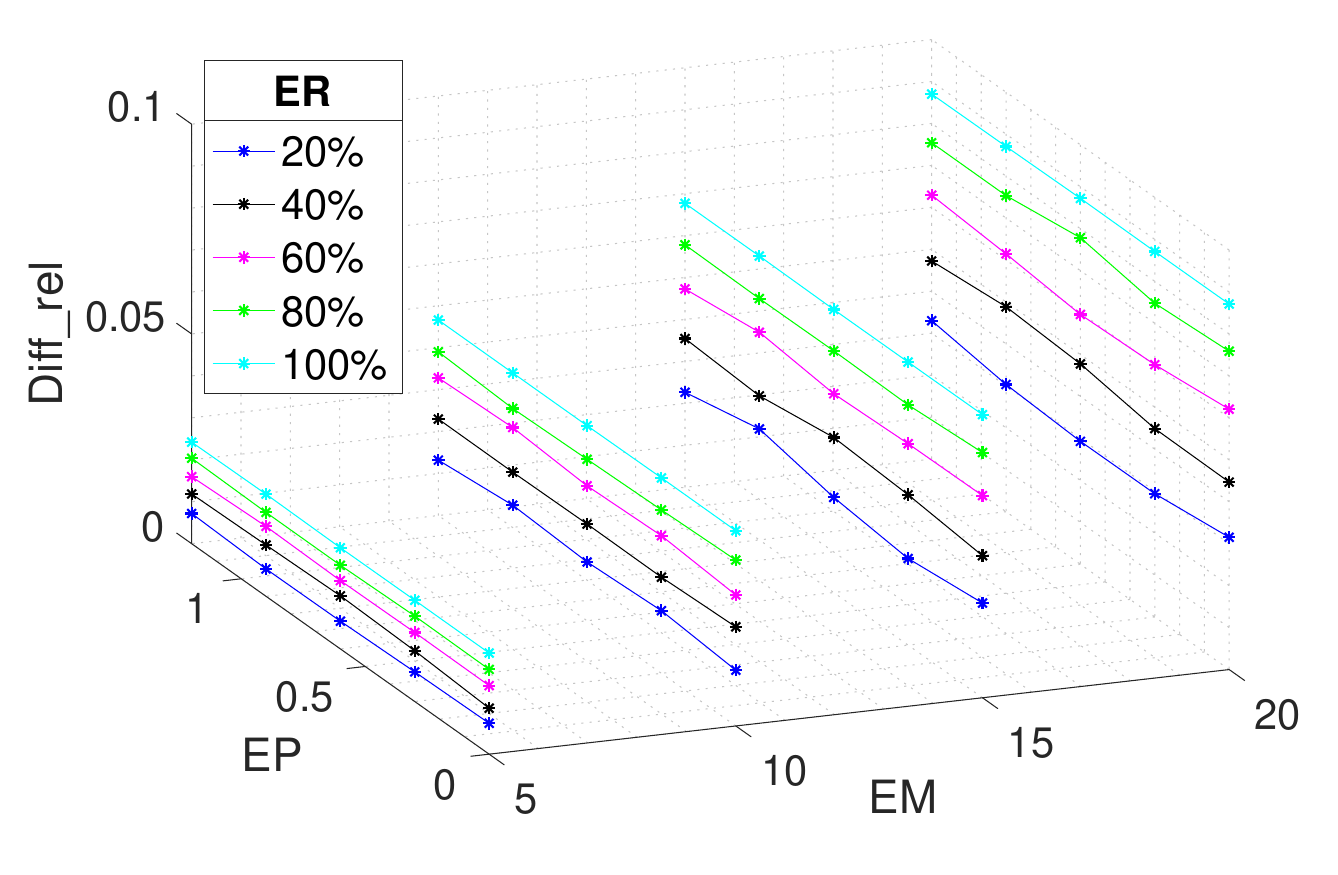}
\caption{StEFCal response for SAM analysis.}
\label{fig:era3}
\end{figure}

\begin{figure}[h!]
     \centering
     \subfloat[][EP$=0$ and EM$=10$.]
     {\includegraphics[width=0.4\textwidth]{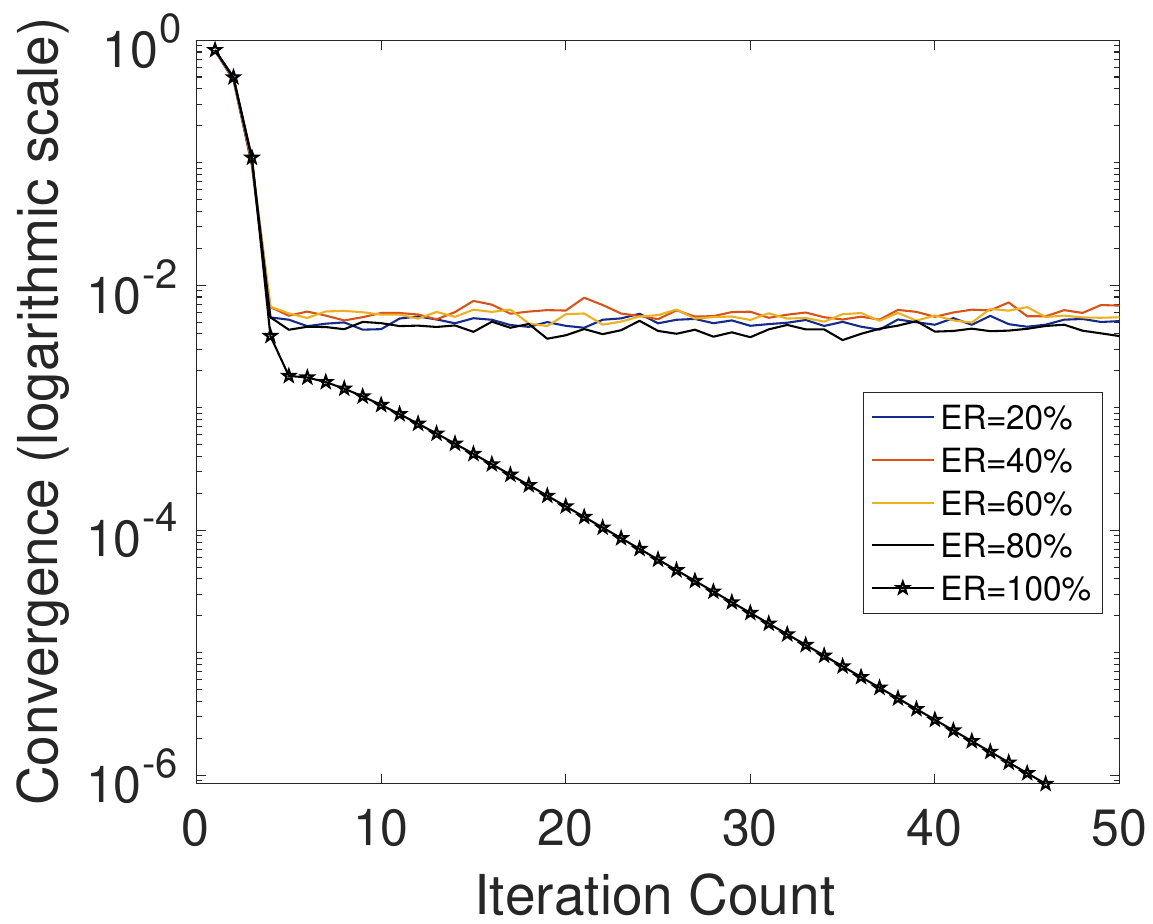} \label{fig:convergence_em10_ep0_ver1}}\hspace{0.25in}
     \subfloat[][EP$=0$ and EM=$20$.]
     {\includegraphics[width=0.4\textwidth]{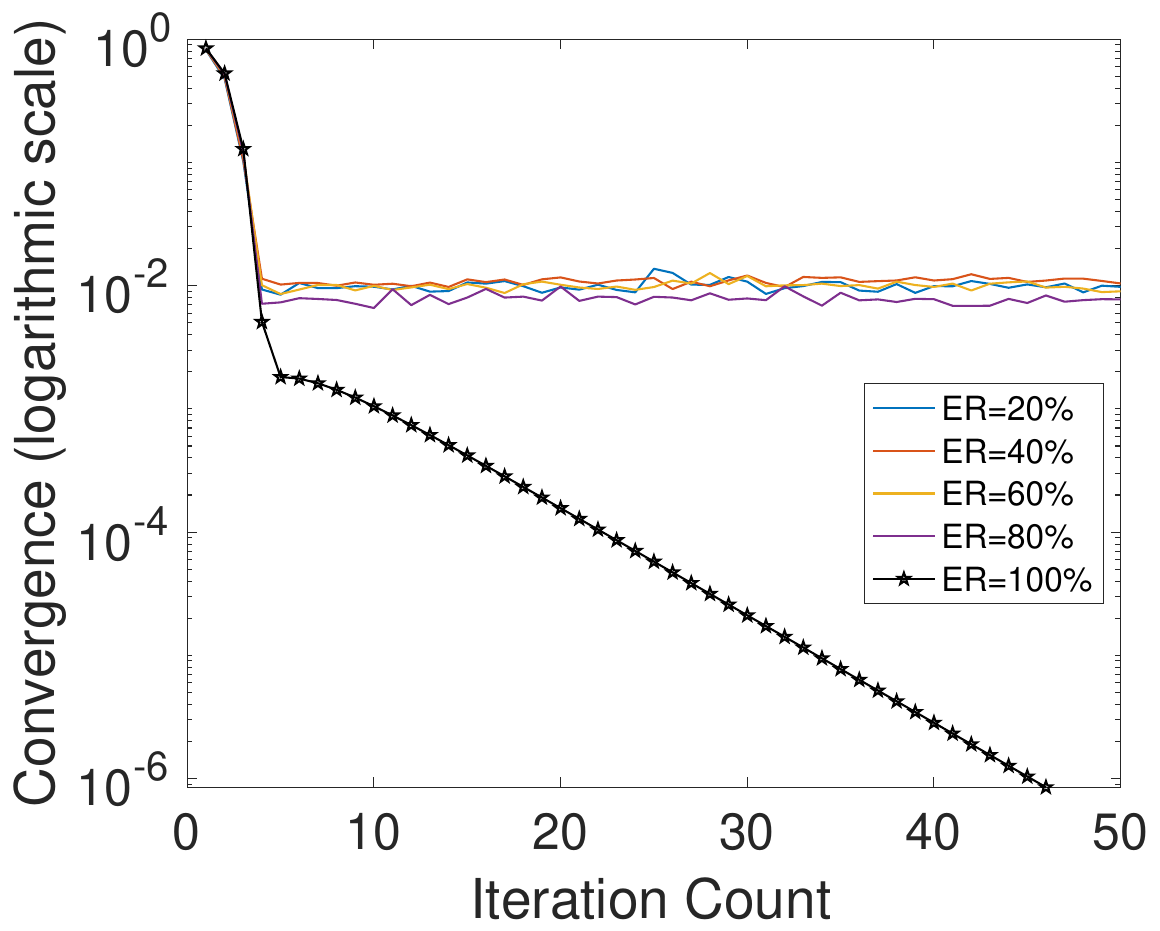} \label{fig:convergence_em20_ep0} }\\
     \subfloat[][EP=$0.1$, EM$=5:5:20$, ER$=20:20:100$.]{\includegraphics[width=0.4\textwidth]{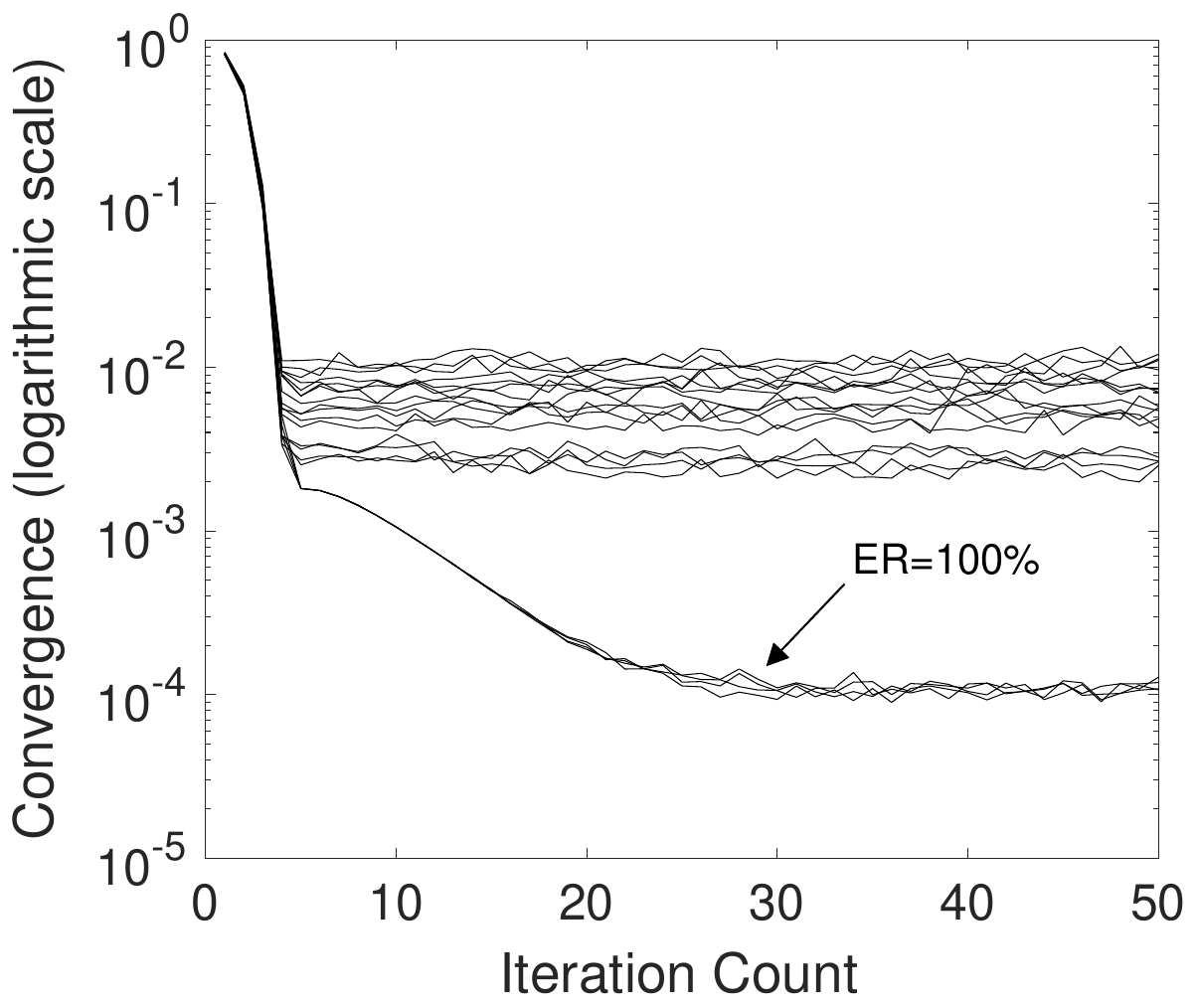} \label{fig:convergence_EM5-5-20_EP0p1_ER20-20-100}}
     \caption{Convergence (logarithmic scale) w.r.t the number of iterations; the algorithm converges at ER$=100\%$ when EP$=0$ (a) and (b). The algorithm does not converge when EP is raised to $0.1$ (c). For the simulations regarding EP$=0.1$ (c), EM has been varied from $5$ to $20$ with a step size of $5$ and ER has been varied from $20\%$ to $100\%$ with a step size of $20\%$.}
     \label{fig:Fig3_convergence}
\end{figure}

Fig. \ref{fig:era3} shows an illustrative example of StEFCal response for SAM analysis. The figure presents the effect of error mean (EM), error predictability (EP) and error rate (ER) on the considered quality metric: Diff\_rel (Eq. \ref{eq:diff_rel}). It can be seen that Diff\_rel is increased as we increase the EM and ER, while it has no remarkable effect due to changes in EP. However, the convergence limit is only achieved at minimum EP and maximum ER, see Fig. \ref{fig:Fig3_convergence}. The iteration count in Fig. \ref{fig:Fig3_convergence} is based on even\footnote{At every even iteration in StEFCal, the gain solution is replaced by the average of the previous (odd) iteration and the current (even) iteration to help fast convergence \cite{107_stefcal}.} iterations, which means that the total iterations are two times the number of iterations shown in the figure. 

\begin{figure}[h!]
\centering
\includegraphics[width=2.8in]{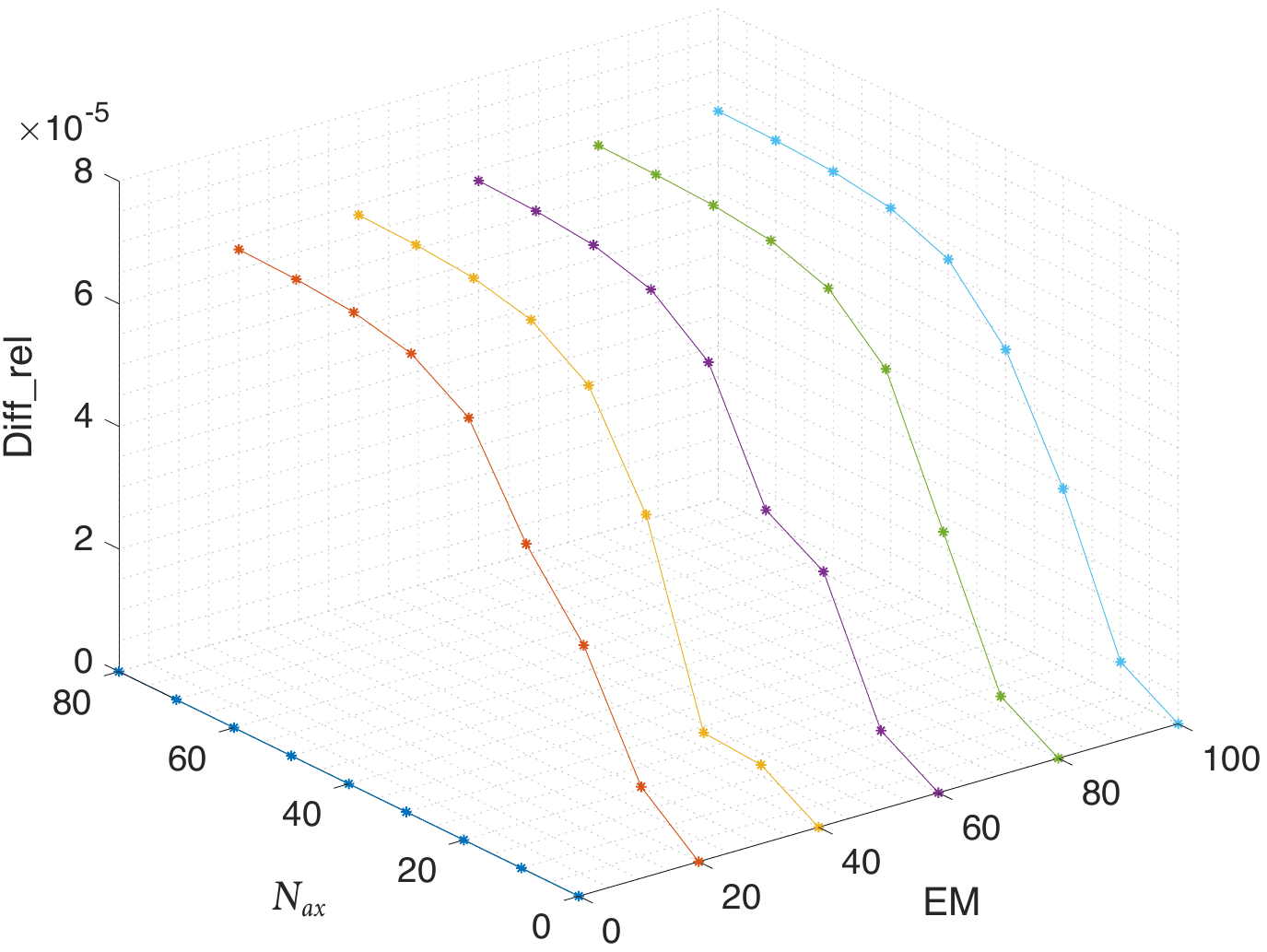}
\caption{StEFCal response for Adaptive-SAM analysis.}
\label{fig:era4}
\end{figure}

Noteworthy, the convergence criterion (Eq. \ref{eq:converg}) does not allow randomly selecting iterations for error injection because this may result in iteration $i-1$ to be computed with an error injection while iteration $i$ to be computed without an error injection. This in turn reduces the precision of the computation among consecutive iterations ($\|\textbf{g}^{[i]} - \textbf{g}^{[i-1]}\|_F$). Accordingly, our SAM analysis simulations show that the convergence is achieved either at ER$=0\%$ (no error injection at all), or at ER$=100\%$ (error injection for all iterations without random skipping of iterations) with suitable EM and EP values, see Fig. \ref{fig:convergence_em10_ep0_ver1} and Fig. \ref{fig:convergence_em20_ep0}.  Likewise, high values of EP also reduce the precision of computation among consecutive iterations, and therefore, do not satisfy the convergence criterion.

In practice, there can be some cases where the benefits of energy efficiency/performance are achieved even if the approximate computation runs for more iterations than the exact computation \cite{104_flops}. However, for the sake of simplicity, we assume that the convergence limit is satisfied when the number of iterations required to converge for error-injected computation is less or equal to that of the exact computing counterpart. To quantify the error resilience intrinsic to StEFCal based on the SAM analysis, extensive simulations have been performed by varying EM, EP, and ER values. The \textit{quality acceptance} has been achieved for EM$\leq0.002$, EP$\leq2.10^{-4}$ at ER$=100\%$. This implies that the algorithm is resilient to an error (Gaussian randomness) that has 
the following maximum values: EM$=0.002$, EP$=2.10^{-4}$,
 and that the error is employed in every iteration. This shows a very small approximation space that does not appreciate employing approximate computing techniques to gain efficiency benefits. 

An illustrative example of Adaptive-SAM analysis of StEFCal is shown in Fig. \ref{fig:era4}. In this case, the simulations are performed for various number of approximate iterations ($N_{ax}$) and error mean (EM) values. However, error predictability and error rate are fixed to the values that allow the solution to converge (EP=$0$ and ER=$100$\%). This means that in every approximate iteration an error of EM$\%$ of the output value is injected. As expected, the Diff\_rel decreases with the decrease in $N_{ax}$, see Fig \ref{fig:era4}. This suggests that the higher the number of exact computing iterations, the better the quality of output and vice versa. 

To quantify the error resilience intrinsic to StEFCal based on the Adaptive-SAM analysis, extensive simulations have been performed by varying $N_{ax}$, EM, EP, and ER values. The \textit{quality acceptance} is achieved for $N_{ax}\leq23\%$, $\text{EM}\leq12$, $\text{EP}\leq0.2$ at $\text{ER}=100\%$. This implies that if the initial (up to) $23\%$ of iterations are approximated, the algorithm is resilient to an error (Gaussian randomness) that has 
the following maximum values: EM$=12$, EP$=0.2$, 
 and that the error is employed in all the initial ($23\%$) iterations. This shows the availability of an approximation space for up to $23$\% of the initial iterations. Therefore, Adaptive-SAM reveals additional error resilience opportunities by quantifying the number of approximate iterations in addition to EM, EP and ER for iterative algorithms. 

 Noteworthy, during the error resilience analysis, the statistical and technique-specific approximation models are applied offline\footnote{Error resilience analysis is performed before the deployment of the application for real use.} and are validated using the quality function\footnote{The quality function is either given within an application/algorithm or has to be defined for error resilience analysis.}. This makes the selection of a quality function very crucial as it decides whether to reject or accept an approximation technique. Therefore, in order to utilize the original quality metric of an iterative application (convergence criterion) in the approximate computing domain, it has to be reconsidered rigorously to attain reliable insights about the approximation space (see Appendix \ref{s:eranalysis:Quality_Function} for an example case study).

\section{Energy Efficient Accelerator Design for Iterative Algorithms}
\label{eeAcceleratorDesign}

So far we have discussed that the statistical approximation models can be applied to evaluate a target algorithm for error resilience. These models inject errors during the execution of the algorithm on statistical bases to quantify the bearable error profile. For iterative applications, our analysis suggests that a certain number of initial iterations can be approximated while producing an acceptable outcome. Therefore, in this section, we present our hardware realization of an accelerator that utilizes a heterogeneous architecture composed of two processing cores\footnote{\textit{Cores} refer to accelerator processing elements within our Least Squares (LS) heterogeneous accelerator architecture. Therefore, the \textit{accurate core} refers to accurate LS accelerator elements and the \textit{approximate core} refers to approximate accelerator elements.}. The two cores differ in their precision of computation, namely an \textit{accurate core} and an \textit{approximate core}. We show how a set of initial iterations can be processed in an approximate core, while the rest of the iterations are processed in an accurate core to achieve an overall energy-efficiency increase.

\subsection{Design of a Heterogeneous Least Squares Accelerator} 
\label{s:eearchitecture:design_accelerator}
\begin{figure}[t!]
\centering
\includegraphics[width=2.4in]{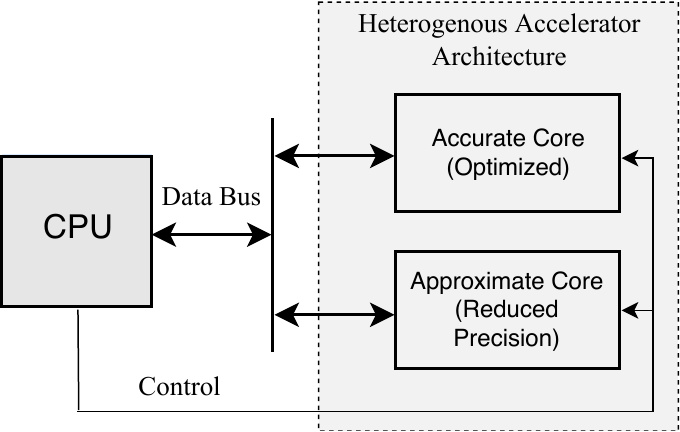}
\caption{Our design methodology for an approximate Least Squares (LS) accelerator enables initial iterations to be processed on an approximate core (while the rest on an accurate core) to achieve an overall energy-efficiency.}
\label{fig:methodology}
\end{figure}
Our design methodology for an approximate Least Squares (LS) accelerator is shown in Fig. \ref{fig:methodology}. The accelerator architecture is composed of two cores that differ in computation precision, introducing heterogeneity in the architecture. The accurate core is optimized for the required precision for the LS algorithm. However, the approximate core introduces a reduced-precision computation to provide energy efficiency. In the proposed LS accelerator, the initial iterations are run on the approximate core, while the rest of the iterations on the accurate core. This brings an overall energy efficiency when a central processing unit (CPU) switches off the unused core. 

Nevertheless, using two cores instead of one brings area overhead. However, if the CPU can utilize both cores simultaneously for parallel processing of independent processes, this area overhead can be translated into increased throughput. In any case, the energy efficiency can be increased for processing the LS algorithm with or without area penalty.

\subsubsection{Overall Energy Savings}
\label{ss:ee:energysavings}
Consider an accurate LS accelerator that utilizes only an accurate core. Let $E_a$ be the total energy consumption of an accurate LS accelerator for processing $N_{acc}$ iterations to solve a given LS problem. Therefore,
\begin{equation}
\label{eq:ee:ea}
E_a= E_{acc} \times N_{acc}
\end{equation}
where $E_{acc}$ is the energy consumption of the accurate core for processing one iteration. Now consider our proposed LS accelerator design that processes $N_{ax}$ iterations utilizing the approximate core while the rest ($N_{acc} - N_{ax}$) is utilizing the accurate core. The energy consumption ($E_h$) of such an accelerator is given as,
\begin{equation}
\label{eq:ee:eh}
E_h= E_{ax} \times N_{ax} + E_{acc} \times (N_{acc} - N_{ax})
\end{equation}  
where $E_{ax}$ refers to the energy consumption of the approximate core for one iteration. For our proposed architecture, we assume that the total number of iterations (running on accurate and approximate cores) remain the same as $N_{acc}$.

The overall energy savings ($S_E$) while utilizing our proposed accelerator design can be given as,
\begin{equation}
\label{eq:ee:es}
S_E= \frac{(E_{a} - E_{h})} {E_{a}}
\end{equation}
Using Eq. \ref{eq:ee:ea} and Eq. \ref{eq:ee:eh}, $S_E$ is given as,
\begin{equation}
\label{eq:es=energy}
S_E= \frac{(E_{acc} - E_{ax}) \times N_{ax}}{E_{acc} \times N_{acc}}
\end{equation}
$E_{acc}$ and $N_{acc}$ are constant terms as they correspond to the reference accurate architecture. On the other hand, reducing $E_{ax}$ or increasing $N_{ax}$ would increase energy benefits ($S_E$). However, lowering $E_{ax}$ means introducing more coarse approximations in the approximate core and this results in practice in a decrease in the number of iterations ($N_{ax}$) that can survive this approximation level. Therefore, there is a trade-off between $E_{ax}$ and $N_{ax}$ and the goal is to find an optimal balance where $S_E$ is maximized. 

\subsubsection{Radio Astronomy Calibration Algorithm (StEFCal)}
We consider a case study of radio astronomy calibration processing that employs a Least Squares (LS) algorithm, StEFCal, as discussed in Section \ref{ss:era:raprocessing}. We demonstrate how to design an LS accelerator using the proposed methodology, wherein the accurate LS core and the approximate LS core are optimized to achieve energy-efficient LS processing. We consider complex data as utilized by StEFCal.  

\begin{figure}[t!]
\centering
\includegraphics[width=3.5in]{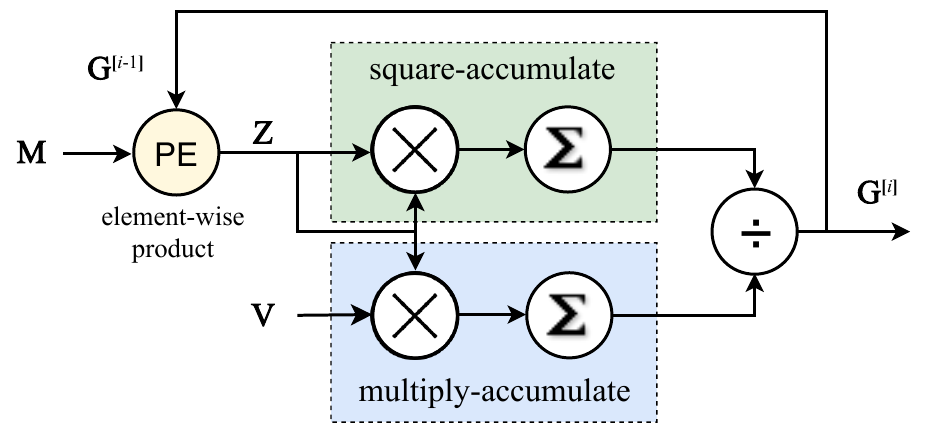}
\caption{Least Squares (LS) algorithm for radio astronomy calibration processing.}
\label{fig:ee:LSalgorithm}
\end{figure}

\begin{figure}
\centering
\includegraphics[width=1\linewidth]{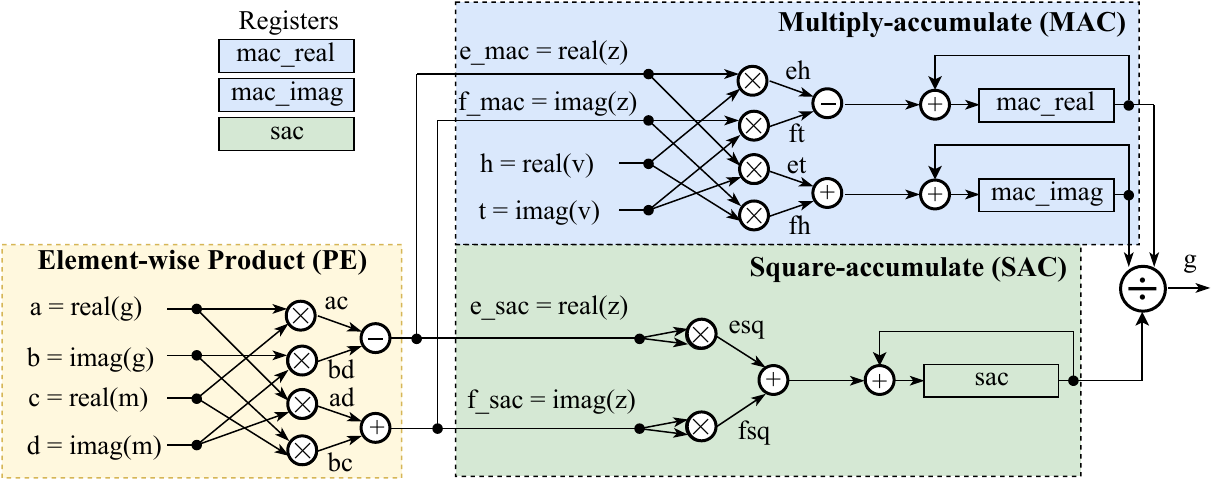}
\caption{Signal flow of least squares algorithm in radio astronomy calibration processing.}
\label{fig:signalflow}
\end{figure}

Fig. \ref{fig:ee:LSalgorithm} shows the four stages of the algorithm. The PE block computes the element-wise product. Square-accumulate (SAC) computes the inner product of $\textbf{Z}$ with itself. Multiply-accumulate (MAC) computes the inner product of $\textbf{V}$ and $\textbf{Z}$. Finally, a division operation is performed to compute the gain values. Considering complex inputs, Fig. \ref{fig:signalflow} illustrates the signal flow of the algorithm. 
It is to be noted that m, v, z and g correspond to the respective elements of $\textbf{M}$, $\textbf{V}$, $\textbf{Z}$, and $\textbf{G}$ matrices.

\subsection{Experimental Results} 
\label{s:eearchitecture:exp}
Here we compare the accurate and the proposed LS accelerator designs as discussed in Section \ref{s:eearchitecture:design_accelerator}. We assume an equal frequency of operation for both designs, which means an equal processing time for executing a single iteration. Therefore, Eq. \eqref{eq:es=energy} is reduced to power ($P$) consumption values only, as in Eq. \eqref{eq:es=power}.
\begin{equation}
\label{eq:es=power}
S_E=\frac{(E_{acc} - E_{ax}) \times N_{ax}}{E_{acc} \times N_{acc}}=\frac{(P_{acc} - P_{ax}) \times N_{ax}}{P_{acc} \times N_{acc}}
\end{equation}

\begin{figure}
\centering
\includegraphics[width=0.98\linewidth]{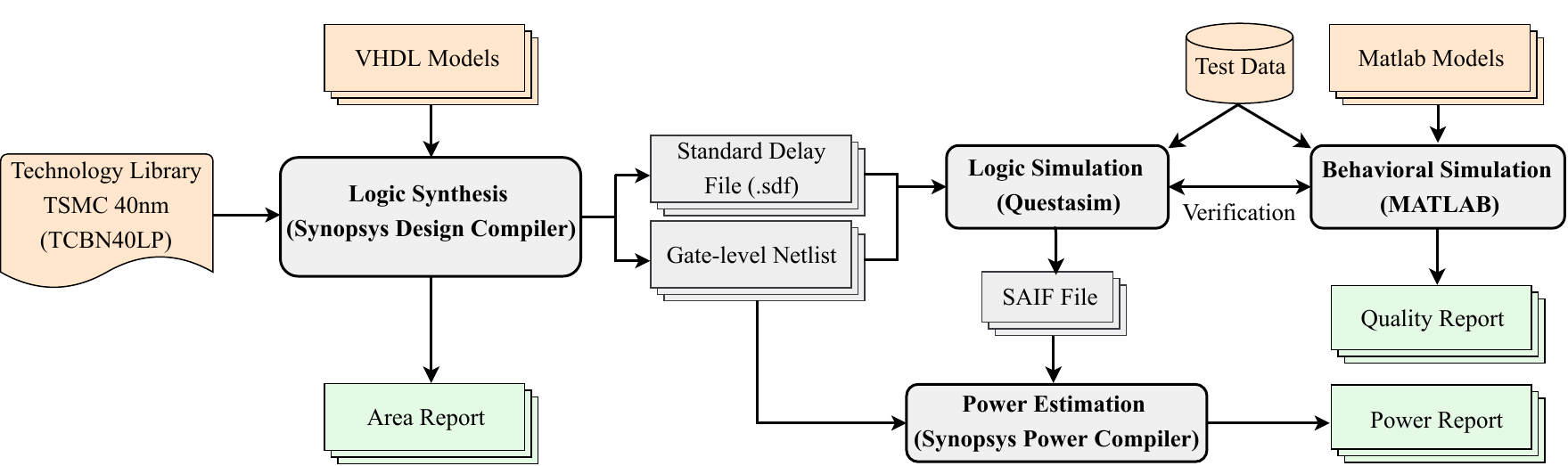}
\caption{Our experimental setup to assess chip-area and power consumption of the considered designs along with their output-quality \cite{exp_setup_squash}.}
\label{fig:ee:experimentalsetup}
\end{figure}

Power consumption and chip-area estimates have been obtained by synthesizing the designs in Synopsys ASIC flow (Design Compiler and Power Compiler) for the TSMC $40$nm Low Power (TCBN$40$LP) technology library at $50$MHz. Our experimental setup is shown in Fig. \ref{fig:ee:experimentalsetup}. Questasim has been utilized to verify the functionality of the synthesized designs (gate-level netlists) and to generate the related SAIF (Switching Activity Interchange Format) files based on the respective standard delay file (.sdf) and test data. The aforesaid SAIF files and gate-level-netlists are utilized by Synopsys Power Compiler for power estimation. 

Output-quality assessment has been performed in Matlab, where the radio astronomy calibration was performed. Here, the Test Data (see Fig. \ref{fig:ee:experimentalsetup}) refers to the radio astronomy calibration data of the LOFAR facility \cite{LOFAR}. Similar to Section \ref{s:eranalysis:Results}, it is assumed that the \textit{quality acceptance} is achieved, if and only if both the convergence (Eq. \ref{eq:converg}) and the Diff\_rel (Eq. \ref{eq:differenceRel}) criteria are satisfied.
\begin{equation}
\label{eq:differenceRel}
  \text{Diff\_rel}=\frac{\|\textbf{g}_{\text{float}}-\textbf{g}_{\text{fixed}}\|_F} {\|\textbf{g}_{\text{float}}\|_F} \leq 10^{-5}
\end{equation}
where $\textbf{g}_{\text{float}}$ and $\textbf{g}_{\text{fixed}}$ refer to the gains obtained using double-precision floating-point and fixed-point computations, respectively.
 
\begin{figure}
     \centering
     \subfloat[][Both computations (float and fixed) converged in 92 iterations.]
     {\includegraphics[width=0.45\textwidth]{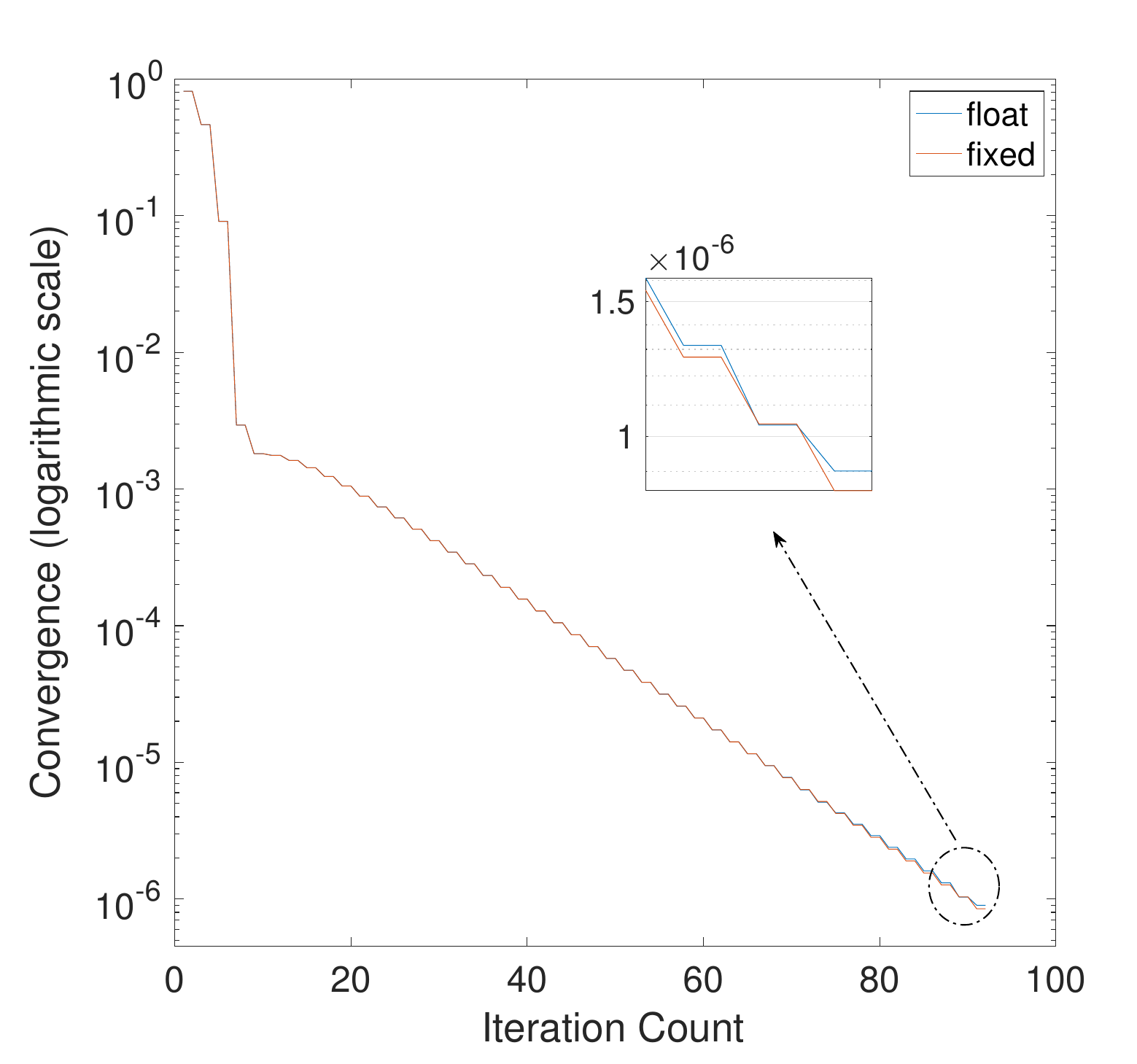} \label{fig:difference:convergence}}\hspace{0.1in}
     \subfloat[width=0.45\textwidth][The minimization of the difference, norm(delta), is achieved for both computations (float and fixed).]
     {\includegraphics[width=0.45\textwidth]{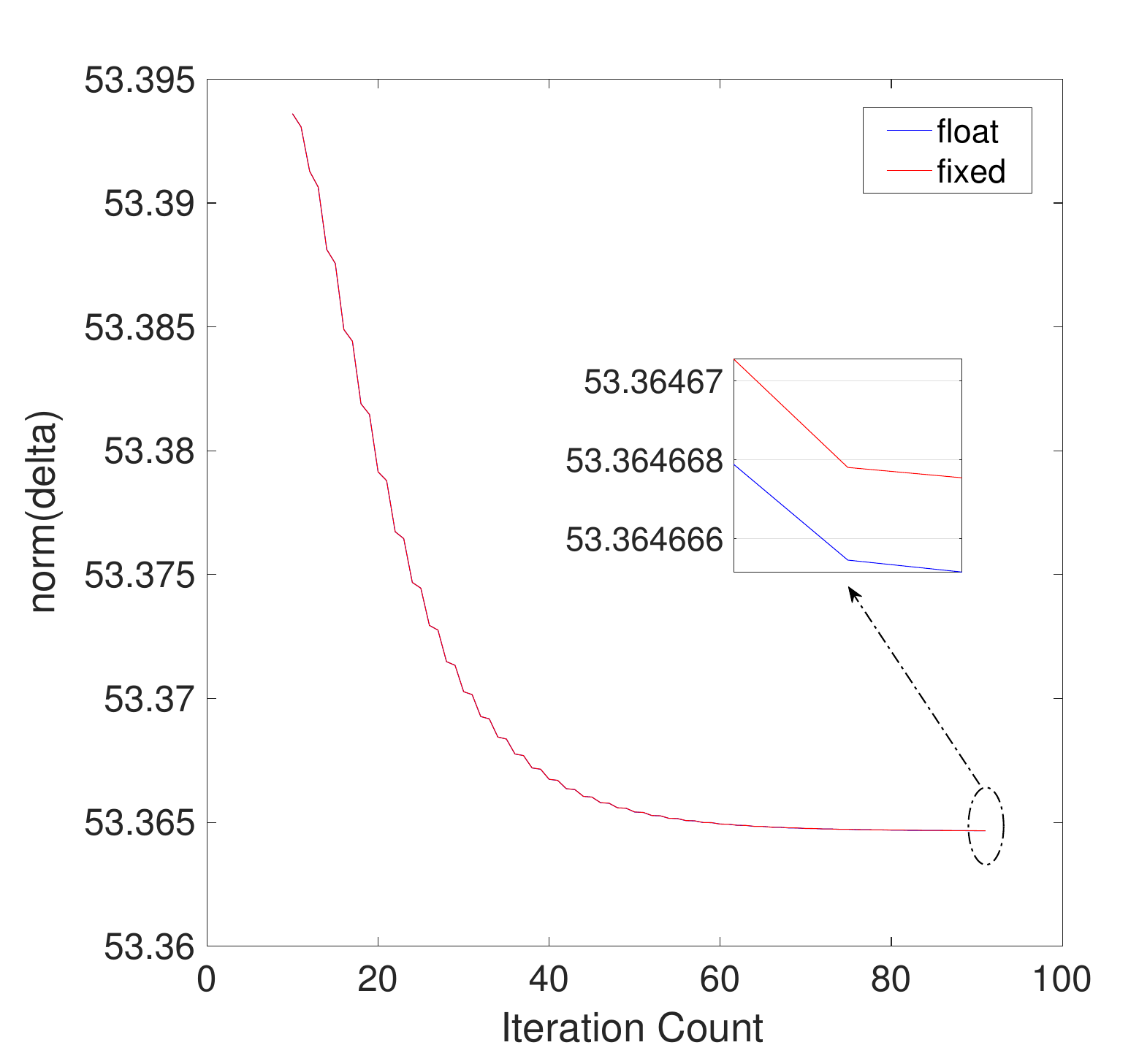} \label{fig:difference:dif}}
     \caption{Comparison between the double-precision floating-point (float) and optimized fixed-point (fixed) StEFCal processing, the latter is referred to as the \textit{accurate core}.}
     \label{fig:difference}
\end{figure}

\begin{figure}
\centering
\includegraphics[width=0.65\linewidth]{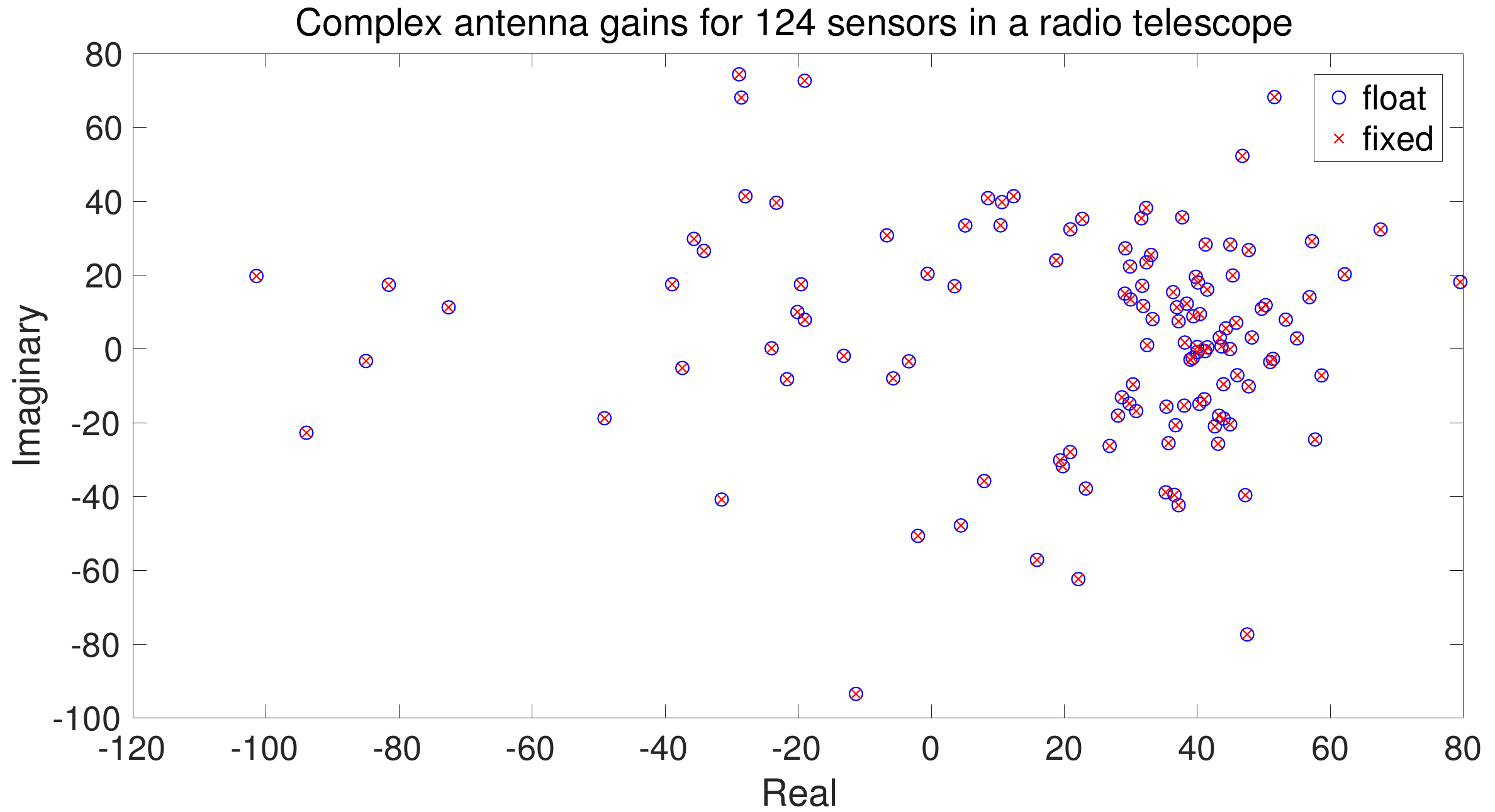}
\caption{Gains computed by double-precision floating-point (float) and optimized fixed-point (fixed) processing.}
\label{fig:gains}
\end{figure}

We assume the accurate core as the optimized fixed-point design of StEFCal, wherein the word length of each signal (shown in Fig. \ref{fig:signalflow}) remains less or equal to $28$ bits. Fig. \ref{fig:difference} shows the comparison between the optimized fixed-point (fixed) and double-precision floating point (float) computation of the StEFCal algorithm. As shown in Fig. \ref{fig:difference:convergence}, both computations converge in $92$ iterations. Fig. \ref {fig:difference:dif} illustrates the behavior of norm(delta)$=\|\textbf{V}-\textbf{GMG}^H\|_F$, which shows that the minimization of the difference is also achieved for the accurate core (fixed). The gain values are shown in Fig. \ref{fig:gains}. As the gain values of the accurate core also satisfy Eq. \ref{eq:differenceRel}, \textit{quality acceptance} is achieved for the accurate core.


As discussed in Section \ref{ss:ee:energysavings}, our proposed LS accelerator design utilizes an approximate core additionally and the overall energy savings can be maximized by optimizing the approximate core in such a way that Eq. \ref{eq:es=power} is maximized. This means that a design space exploration is to be carried out to find a trade-off point where $P_{ax}$ and $N_{ax}$ values maximize Eq. \ref{eq:es=power}. We consider truncation of inputs as a means of approximation in the approximate core, wherein four multipliers of the MAC and two squarers of the SAC have been approximated (see Fig. \ref{fig:signalflow}). Based on our design space exploration, the number of truncated bits to obtain the optimized approximate core are shown in Table \ref{tab:ee:widths}, where the resulted $N_{ax}=52$. The corresponding input widths of the accurate core are also shown as a reference.

\begin{table}[t!]
\caption{Bit widths of the StEFCal algorithm for accurate LS core and the truncation levels to achieve an optimal approximate LS core.}
\label{tab:ee:widths}
\centering
\begin{tabular}{p{50pt}| p{67pt}| p{77pt}}
\bfseries Signal name & \bfseries Accurate Core (bit-width) & \bfseries Approximate Core (bits truncated) \\ \hline 
\textbf{h} & 18 & 0 \\ \hline
\textbf{t} & 18 & 0 \\ \hline
\textbf{e\_sac} & 21 & 8 \\ \hline
\textbf{f\_sac} & 20 & 8 \\ \hline
\textbf{e\_mac} & 23 & 8 \\ \hline
\textbf{f\_mac} & 24 & 12 \\ \hline
\end{tabular}
\end{table}
\begin{figure}[t!]
\centering
\includegraphics[width=0.7\linewidth]{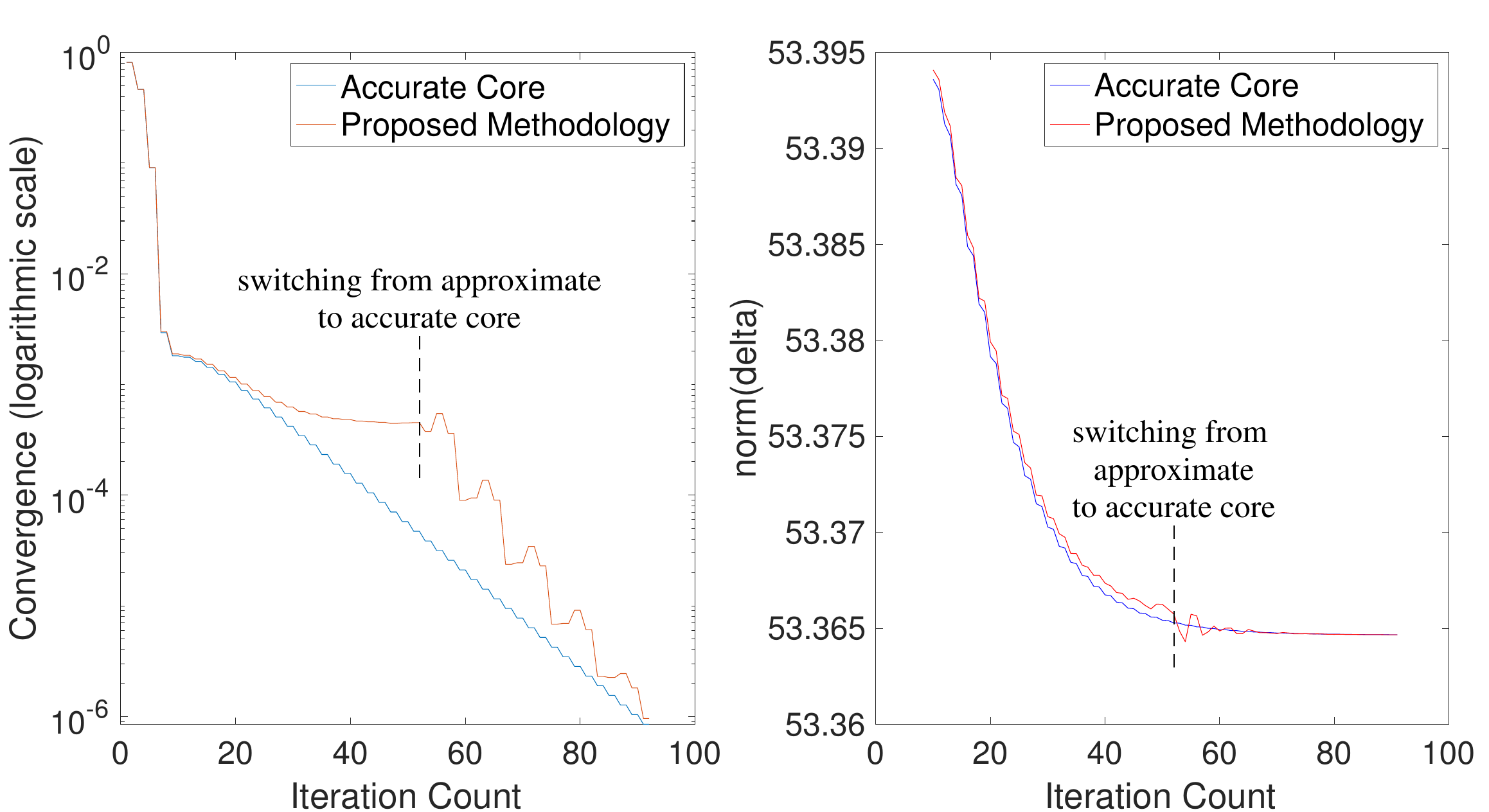}
\caption{StEFCal processing based on our proposed heterogeneous methodology as compared to that of an accurate design. Our proposed methodology processes the first $52$ iterations on the approximate core and the rest on the accurate core.}
\label{fig:tinp_results}
\end{figure}
Fig. \ref{fig:tinp_results} shows the output-quality comparison of our proposed methodology with that of the accurate core. The \textit{quality acceptance} is achieved as we process the first $52$ iterations on the approximate core, and the rest of the iterations ($40$) on the accurate core. After switching to the accurate core, two phenomena can be noticed: (1) the precision-oriented metric, i.e., convergence, experiences jumps because of the increase (change) in computation precision, see Fig. \ref{fig:tinp_results}-left. (2) the deviation from the accurate solution gradually decreases, see Fig. \ref{fig:tinp_results}-right. Overall, the solution converges to an acceptable value in the same number of iterations, i.e., $92$. 

\subsubsection{Energy Efficiency Improvements}

Table \ref{tab:ee:areapower} shows the chip-area and power consumption of the accurate and approximate cores. It can be seen that the approximate core offers $41$\% power reduction as compared to that of accurate core ($P_{ax}=2.08mW$ and $P_{acc}=3.55mW$). As $N_{ax}=52$ and $N_{acc}=92$, the overall energy saving using Eq. \ref{eq:es=power} for our proposed LS accelerator design is $23.4$\%. The aforesaid energy saving is obtained while comparing the accurate single-core design with our proposed two-core design assuming that only one core is switched on at a certain period of time.

\begin{table}[t!]
\caption{Area and Power comparison of the accurate and approximate LS cores for TSMC $40$nm Low Power (TCBN$40$LP) technology.}
\label{tab:ee:areapower}
\centering
\begin{tabular}{p{78pt}| p{57pt}| p{58pt}}
\bfseries LS Core Type & \bfseries Area ($\mu m^2$) & \bfseries Power ($mW$) \\ \hline 
Accurate Core & 27023 & 3.55 \\ \hline
Approximate Core & 20604 & 2.08 \\ \hline
\end{tabular}
\end{table}

\subsubsection{Discussion and Future work} 
\label{DandF}
While comparing the chip-area of our proposed accelerator design (a two-core architecture) with that of an accurate accelerator design (a single-core architecture), the area overhead is $76$\%. This area overhead is due to the addition of the approximate core. However, if a CPU (see Fig. \ref{fig:methodology}) can utilize both cores simultaneously for independent processes, the area overhead can be translated into throughput increase while having the same energy efficiency benefits for each process.     

A case study of radio astronomy calibration (StEFCal) processing has been discussed to show how to employ the proposed heterogeneous accelerator by using a single time slot of LOFAR \cite{LOFAR} data. Nevertheless, an increased data set would better provide the allowable number of iterations to run on an approximate core, therefore, a better estimate of energy savings that can be achieved using the proposed accelerator architecture. Moreover, our architecture does not affect the stability of the iterative process as the approximate (relatively low-precision) core is utilized for initial iterations when (relativey) high-precision is not required, see Appendix \ref{appendix:stability} for stability discussion.

\section{Conclusions}
We have addressed energy-efficient processing of iterative workloads from its concept to implementation. We exploited the error resilience of the workloads and presented our results with a case study of radio astronomy processing (Least Squares acceleration). Our proposed adaptive statistical approximation model (Adaptive-SAM) has shown improvements in the error resilience analysis of iterative algorithms. It quantifies the number of resilient iterations in addition to statistical analysis parameters. Moreover, we have shown that the quality function must be reconsidered in the error resilience analysis process as the original quality metric of an iterative algorithm (convergence criterion) might not be necessarily sufficient. In which case, an additional quality metric has to be defined for reliable quality assessment to establish the promising approximate computing strategies.

A heterogeneous architecture for Least Squares (LS) acceleration has been presented targeting energy-efficiency. We have shown how a combination of optimized-precision (accurate) and reduced-precision (approximate) computing cores can be utilized to provide acceptable quality output while reducing energy consumption as compared to that of an accurate optimized architecture. Our design methodology exploits the inherent error-resilience of an iterative workload to leverage an approximate computing core for processing the initial iterations of the LS algorithm. A case study of radio astronomy calibration processing has shown $23\%$ of energy savings as compared to that of the accurate counterpart. However, it is to be noted that the proposed methodology is independent of the application, provided that the computation pattern is iterative in nature. We have utilized input truncation as the means of approximations within an approximate core. However, our methodology can utilize any approximation technique that is promising for the target application.

\begin{acks}
This research was conducted in the context of the ASTRON and IBM joint project, DOME, funded by the Netherlands Organization for Scientific Research (NWO), the Dutch Ministry of EL\&I, and the Province of Drenthe. The authors are thankful to Dr. Faizan Ahmed (FMT Group, University of Twente) for his suggestions about the stability of the iterative algorithms and his review of Appendix B.
\end{acks}

\bibliographystyle{ACM-Reference-Format}
\bibliography{sample-base}

\appendix

\section{Significance of Quality Function Reconsideration} 
\label{s:eranalysis:Quality_Function}

In our case study of an iterative algorithm (StEFCal), the convergence criterion is utilized for the exact computing case. It computes the relative distance in Euclidean space between the current and previous solution vectors. The convergence is assumed to be satisfied if the improvement within two consecutive iterations is less or equal to $1.10^{-6}$, which means that the solution has already been converged and no further precision can be achieved by computing more iterations. However, in the error resilience analysis process, it cannot be guaranteed that the acceptable solution is achieved when it satisfies convergence. Perhaps, the solution is converged in the wrong plane, which is very distant from that of acceptable solution plane. 

\begin{figure}
     \centering
     \subfloat[][Case 1: Convergence.]
     {\includegraphics[width=0.4\textwidth]{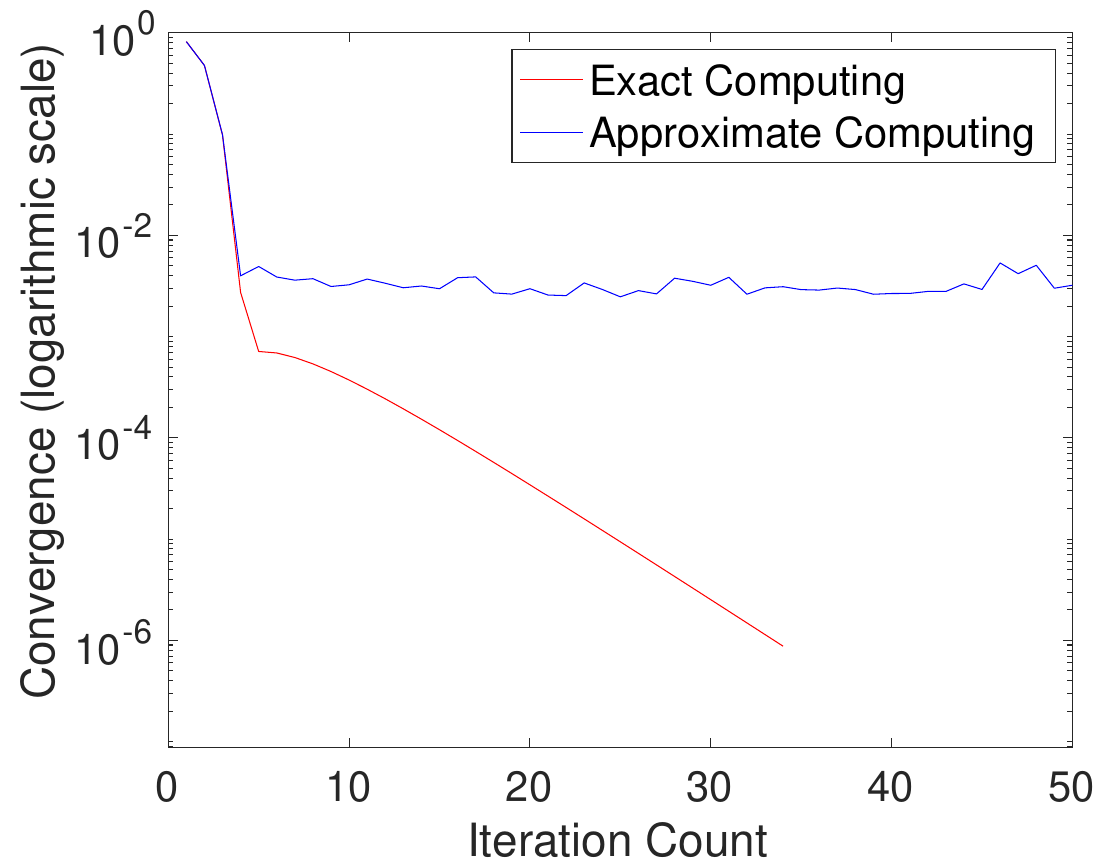} \label{fig:era5a}}\hspace{0.25in}
     \subfloat[][Case 1: Complex sensor gains ($P=124$).]
     {\includegraphics[width=0.4\textwidth]{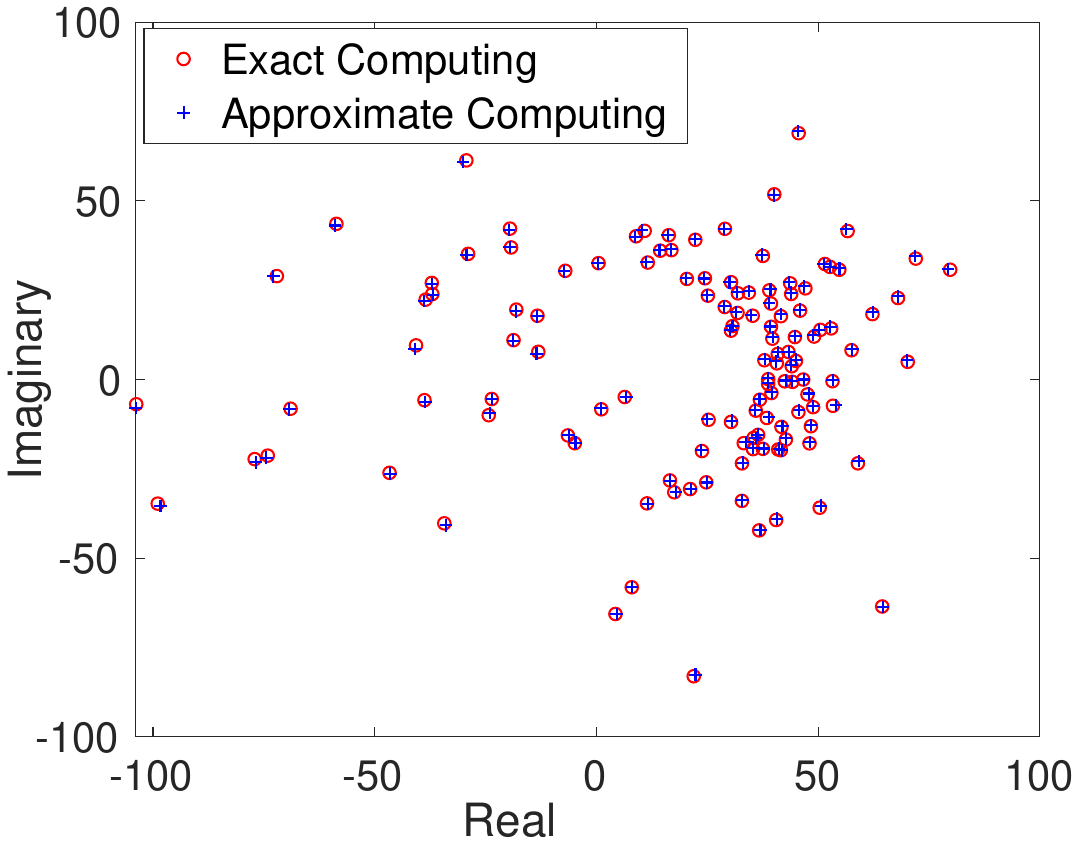} \label{fig:era5b}}\\
     \subfloat[][Case 2: Convergence.]{\includegraphics[width=0.39\textwidth]{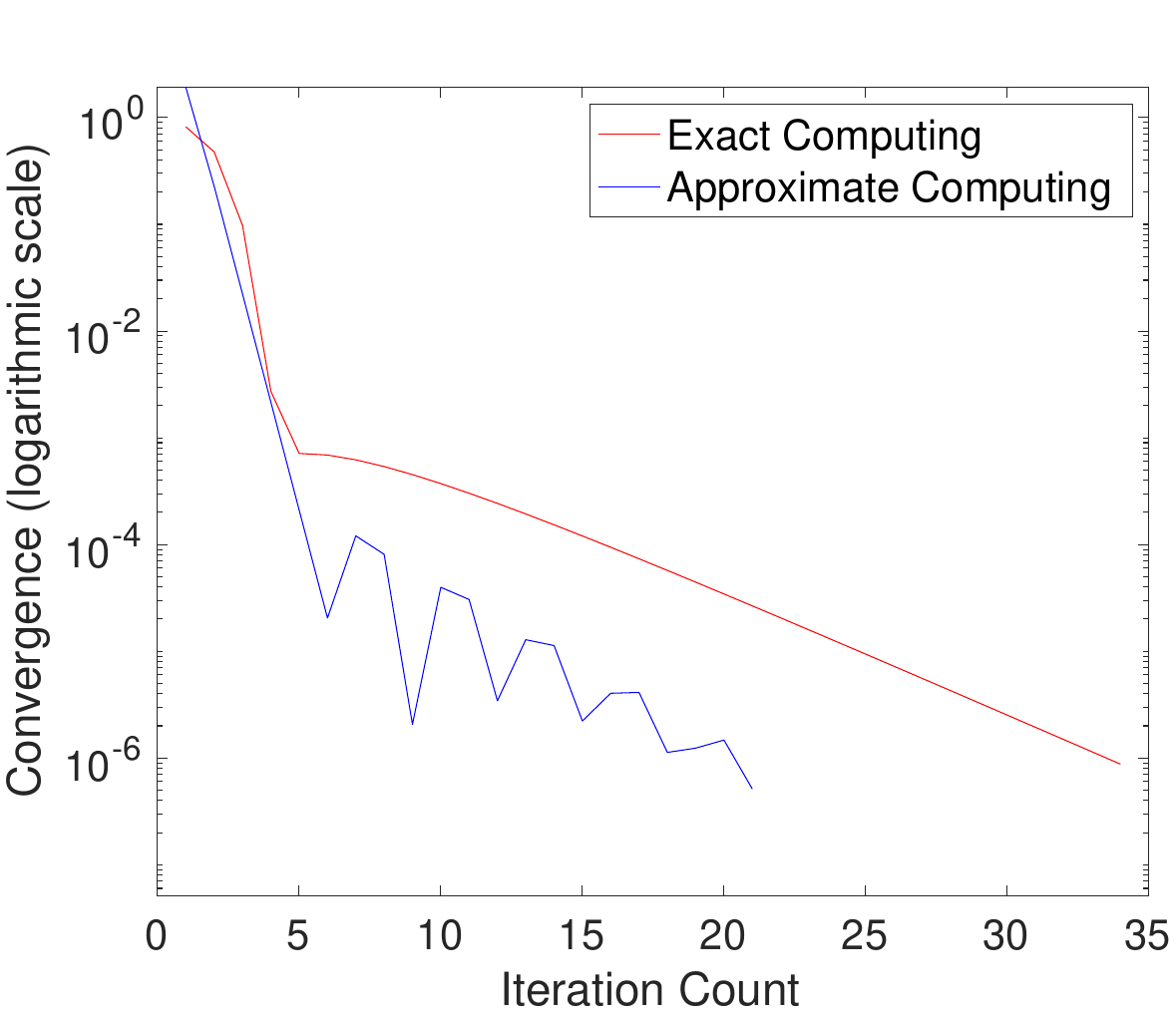} \label{fig:era5c}}\hspace{0.25in}
     \subfloat[][Case 2: Complex sensor gains ($P=124$).]{\includegraphics[width=0.41\textwidth]{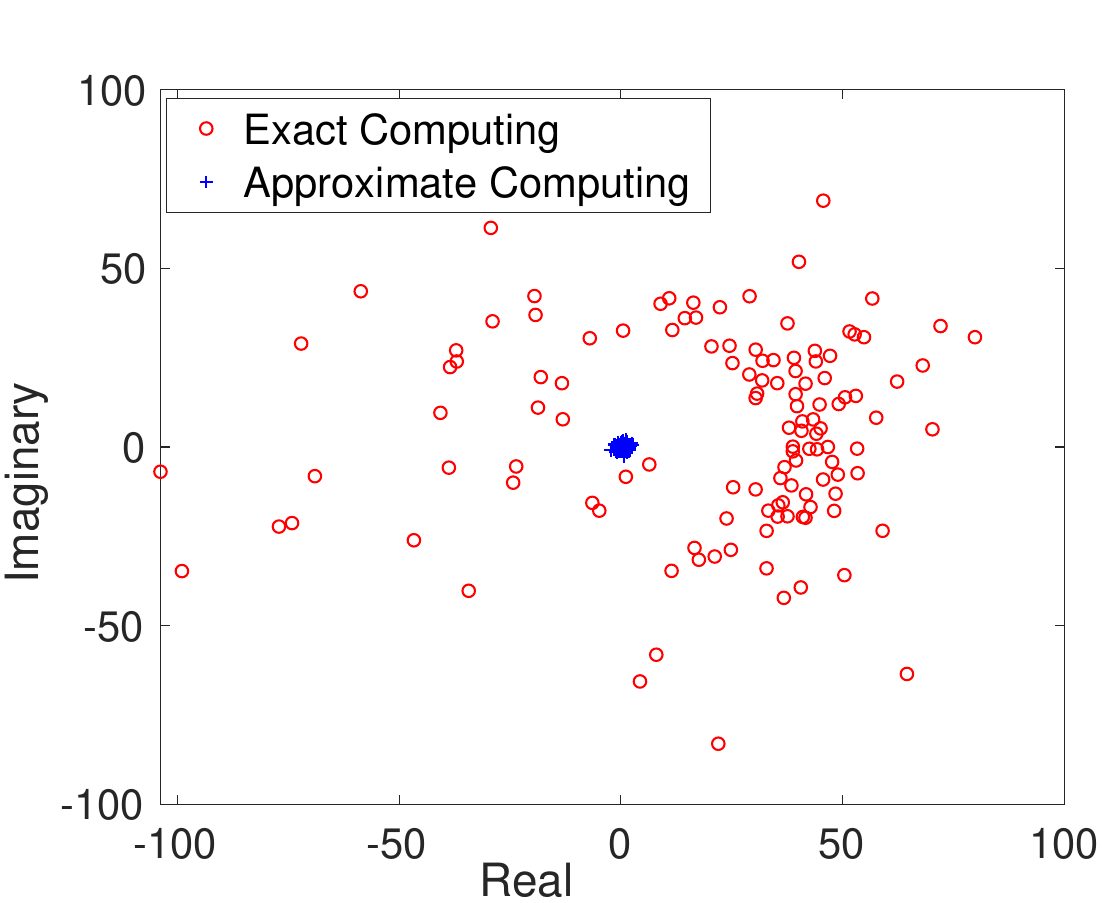} \label{fig:era5d}}
     \caption{Error resilience identification; convergence (logarthmic scale) w.r.t the number of iterations (a) and (c); complex sensor gains for $P=124$ (b) and (d). Case 1 and Case 2 represent a random skipping of a computation and an arbitrary error injection, respectively, in each iteration. Similar to Fig. \ref{fig:Fig3_convergence} the iteration count is based on even iterations.}
     \label{fig:era5}
\end{figure}

We have observed this phenomenon during the error-resilience identification phase. As discussed in Section \ref{s:eranalysis:relatedwork}, random errors are injected into the resilience identification analysis while using a relaxed quality function. Two cases are shown in Fig. \ref{fig:era5} to illustrate the problem. The first case is randomly skipping computations (Fig. \ref{fig:era5a} and Fig. \ref{fig:era5b}), while the second case is an arbitrary error injection in the first element of each column of the $\textbf{Z}$, i.e., $\textbf{Z}_{:,p}$ (Fig. \ref{fig:era5c} and Fig. \ref{fig:era5d}). 

For the first case, we can see a comparable response of approximate and exact computing for gains outputs, albeit with no quality improvement beyond a specific number of iterations for the convergence metric. However in the second case, although the approximate solution converges quickly (Fig. \ref{fig:era5c}), it produces unacceptable gains (Fig. \ref{fig:era5d}). This shows that the original quality metric of StEFCal (the convergence criterion) is not sufficient in the resilience analysis process. For that matter, we introduced an additional quality parameter: Diff\_rel (Eq. \ref{eq:diff_rel}) that can ensure the convergence of an approximate solution within an acceptable distance to that of an exact solution in the Euclidean space. Therefore in Section \ref{s:eranalysis:Results}, we assume that the \textit{quality acceptance} is achieved if and only if both the convergence and Diff\_rel criteria are satisfied.

\section{Stability of Iterative Process}
\label{appendix:stability}
As given in \cite{107_stefcal}, this section discusses the stability of the iterative  algorithm that is utilized in our case-study. We reproduce the results as given in \cite{107_stefcal} for the sake of completeness. The scalar measurement equations or data model are used for calibrating radio astronomical system. The StEFCal is used to estimate the complex antenna gains $g_p$ for the $P$ sensors in radio telescope \cite{107_stefcal}. It is shown in \cite{107_stefcal} that the receiver path gains can be computed by solving the following optimization problem
\begin{equation}\label{optProblem}
\hat{\textbf{g}}=argmin_{\textbf{g}} \left\|\hat{R}-GMG^H\right\|^2_2
\end{equation}
where $\hat{R}$ is the array covariance matrix, $G=diag(\textbf{g})$ is a diagonal matrix whose diagonal are the vector of complex valued receiver path gain $\textbf{g}$, while $M$ is the model convariance matrix of the observed scene. It is clear that the variable $G$ cannot be separated in the equation (\ref{optProblem}). To solve the problem we use implicit scheme in which once we fix $G$ and solve for $G^H$ and vice versa. In this case the iteration function simplifies to
\begin{equation}\label{optProblem_2}
G^{[i]}=argmin_{G} \left\|\hat{R}-G^{[i-1]}MG^H\right\|^2_2
\end{equation}
Introducing, $Z^{[i]}=G^{[i]}M$, we get
\[\left \|\hat{R}-ZG^H\right \|=\sqrt{\sum_{p=1}^{P}{\left\|\hat{R}_{:,p}-Z_{:,p}\textbf{g}_p\right\|_2^2}}\]
Note that the above is a system of $P$ independent least square problem which can be solved by using the following iteration:
\begin{equation}\label{eq:mainIteration}
g_p^{[i]} =\frac{R^H_{:,p}\cdot Z^{[i-1]}_{:,p}}{\left(Z^{[i-1]}_{:,p}\right)^H\cdot Z^{[i-1]}_{:,p}}
\end{equation}
where $R^H_{:,p}$ is the Hermitian transpose of array covariance matrix's $p^{th}$ column, and $Z^{[i-1]}$ is the element-wise product of $g_p^{[i-1]} $ and the model covariance matrix's $p^{th}$ column. 

\subsection{Convergence}
Let us assume that the array covariance matrix $\hat{R}$ is exactly represented by the model covariance matrix that substitute $\hat{R}$ with $G^HMG=gg^H\odot M$. By noting that $\hat{R}_{:,p}=[gg^H\odot M]_{:,p}= (\textbf{g} \odot M_{:,p}) g_p$, we have

\begin{equation}\label{eq:mainIteration_2}
g_p^{[i]} =\frac{\left(\textbf{g} \odot M_{:,p}\right)^H \left(g^{[i-1]}\cdot M_{:,p}\right)}{\left(g^{[i-1]}\cdot M_{:,p}\right)^H\left(g^{[i-1]}\cdot M_{:,p}\right)}g_p
\end{equation}
Let $\mathbf{a},\mathbf{b}$ be two vectors, the weighted inner product with respect to weight vector $\mathbf{w}$ can be written as 
\[<a,b>_w=\sum_i{w_i a_ib_i}.\]
Recognizing that $w_p=M^*_{:,p}\odot M_{:,p}$ can be used as a weight for the inner product than the equation \ref{eq:mainIteration_2}, reduces to
\begin{equation}\label{eq:mainIteration_3}
g_p^{[i]} =\frac{\langle \textbf{g} ^{[i-1]},\textbf{g}\rangle _{w_p}}{\langle \textbf{g} ^{[i-1]},\textbf{g}^{[i-1]}\rangle _{w_p}}g_p
\end{equation}
Let $\alpha$ be a scaling factor and $\epsilon$ is the error vector than the initial estimate can be written as $\textbf{g}^{[i]}=\alpha(\textbf{g}+\epsilon)$, reducing the iteration (\ref{eq:mainIteration_3}) to 
\begin{equation}\label{eq:mainIteration_3}
g_p^{[i]} =\frac{1}{\overline{\alpha}}\frac{\langle \textbf{g}+\epsilon,\textbf{g}\rangle _{w_p}}{\|\textbf{g}+ \epsilon\|^2 _{w_p}}g_p=\frac{1}{\overline{\alpha}} \beta_p g_p
\end{equation}
where $\beta_p=\frac{\langle \textbf{g}+\epsilon,\textbf{g}\rangle _{w_p}}{\|\textbf{g} + \epsilon\|^2 _{w_p}}$. Notice that the full iteration is the average of two iterations. Therefore, the full iteration can be written as,
\[
\textbf{g}^{[i+1]}=\frac{1}{2}\left(\alpha \Tilde{\beta}\odot \textbf{g}+\frac{1}{\overline{\alpha}}\odot \textbf{g} \}\right) =\frac{1}{2} \left(\frac{|\alpha|\Tilde{\beta}+\beta}{\overline{\alpha}}\right)\odot \textbf{g}
\]
where $\beta=[\beta_1,\cdots, \beta_P]^T$ and $\Tilde{\beta}_p=\langle \beta \odot \textbf{g} ,\textbf{g}\rangle _{w_p}/\|\beta\odot \textbf{g}\|^2_{w_p}$.

The convergence of iterative methods is not global in general. To derive conditions under which the above iteration converges, consider two initial estimates $g_1^{[0]}$ and $g_2^{[0]}$ and we like to show that:  
\begin{equation}\label{ineq:convergence}
\left \| g_1^{[2]}-g_2^{[2]}\right \| \le \left \| g_1^{[0]}-g_2^{[0]}\right\|
\end{equation}
 Since we are averaging in even-odd iteration in StEFCal algorithm, therefore, we have considered $g_1^{[2]}$ as a next iteration instead of $g_1^{[1]}$. The above equation simply states that for two different starting points the difference between two consecutive values of the iteration must be bounded above by the actual difference in the initial guesses. To further simplify the above inequality we perform a substitution. 
Consider again  that the initial vector is scaling of error and the true gain i.e., $\textbf{g}^{0}_1=\alpha_1 (\textbf{g}+\epsilon)$ and $\textbf{g}^{0}_2=\alpha_2 (\textbf{g})$. This reduces the inequality (\ref{ineq:convergence}) to\color{black} 
\begin{equation}\label{ineq:cong_2}
    \left \| \frac{1}{2} \left(\frac{|\alpha|\Tilde{\beta_1}+\beta_1}{\overline{\alpha_1}}- \frac{|\alpha|\Tilde{\beta_2}+\beta_2}{\overline{\alpha_2}}\right)\odot \textbf{g} \right\| \le \|\alpha_1 (\textbf{g}+\epsilon)-\alpha_2 \textbf{g}\|
\end{equation}
\noindent From $\beta_p=\frac{\langle \textbf{g}+\epsilon,\textbf{g}\rangle _{w_p}}{\|\textbf{g} + \epsilon\|^2 _{w_p}}$ it is clear that if $\epsilon =0$
 than $\beta=\textbf{1}$ and consequently $\Tilde{\beta}=\textbf{1}$, where $\textbf{1}$ is the vector of ones. Thus the inequality (\ref{ineq:cong_2}) reduces to,
 \begin{align*}
 \left \| \frac{1}{2} \left(\frac{\overline{\alpha_2}|\alpha|\Tilde{\beta_1}+\overline{\alpha_2}\beta_1-\overline{\alpha_1}|\alpha_2|\textbf{1}-\overline{\alpha_1}\textbf{1}}{\overline{\alpha_1}~~\overline{\alpha_2}}\right)\odot \textbf{g} \right\| 
 &=\frac{1}{2} \left \|  diag\left(\alpha_1 \Tilde{\beta_1}-\alpha_2 \textbf{1} +\frac{\overline{\alpha_2}\beta_1 -\overline{\alpha_1}\textbf{1}\}}{\overline{\alpha_1}~~\overline{\alpha_2}}\right) \textbf{g} \right\|\\
 &\le \frac{1}{2} \left \|  diag\left(\alpha_1 \Tilde{\beta_1}-\alpha_2 \textbf{1} +\frac{\overline{\alpha_2}\beta_1 -\overline{\alpha_1}\textbf{1}\}}{\overline{\alpha_1}~~\overline{\alpha_2}}\right) \right\| \left\| \textbf{g}\right\| \\
 &=\frac{1}{2} \sigma_{max}\left(\left \|  diag\left(\alpha_1 \Tilde{\beta_1}-\alpha_2 \textbf{1} +\frac{\overline{\alpha_2}\beta_1 -\overline{\alpha_1}\textbf{1}\}}{\overline{\alpha_1}~~\overline{\alpha_2}}\right) \right\|\right) \left\| \textbf{g} \right\|
\end{align*}
In the above, we used the definition that for a matrix $A$ we have $\|A\|=\sigma _{max}(A)$, where $\sigma _{max}$ is the maximum singular value of the matrix $A$. Note also that for a diagonal matrix, the maximum singular value is in fact the maximum value at its diagonal. Denoting the $p$ for which the the maximum is achieved by $p _{max}$ we have,

\[\frac{1}{4}\left |\alpha_1 \Tilde{\beta}_{1,p_{max}}-\alpha_2 +\frac{\overline{\alpha_2}\beta_{1,p_{max}}-\overline{\alpha_1}}{\overline{\alpha_1}~~\overline{\alpha_2}}\right |^2 \le |\overline{\alpha_1}-\overline{\alpha_2}|^2+|\alpha_1|^2 \frac{\|\epsilon\|^2}{\|\textbf{g}\|^2}\]
the above inequality represents the condition for convergence for the iteration given in (\ref{eq:mainIteration}).

The convergence of the StEFCal is proven by showing that the iteration (\ref{eq:mainIteration}) is indeed a contraction mapping \cite{107_stefcal}. 
However, the convergence is not global, meaning that the initial guess needs to satisfy certain conditions for the algorithm to converge to a solution. The most important condition of convergence is that the initial estimate should not be orthogonal to the true value of the solution, i.e., the inner product between the initial guess and the true solution should be nonzero. Moreover, the initial estimate must not be close to zero.

\par During the offline analysis, applying SAM and Adaptive-SAM raise further questions on the convergence of the algorithm. Before describing the application, we describe what is known as the stability of the iteration.  As described above, the algorithm is not globally convergent; therefore, it becomes important to know the magnitude of change in the solution as produced by the algorithm with respect to the change in the initial estimate. In our case, this is more important since the exact core uses the answer as produced by the approximate core as the initial guess. To ensure stability, one needs to introduce an extra check during the offline analysis, where the output of the approximate core is tested to ensure that it does not lie in the null space of the true solution. If the output does not lie in the null space of the true solution, it is safe to use the output as an initial guess in the exact core. 

\end{document}